\documentclass[twocolumn,twoside]{IEEEtran}
\usepackage{comment}
\usepackage{xcolor,soul,framed} 
\usepackage{amsmath}
\colorlet{shadecolor}{yellow}
\usepackage[pdftex]{graphicx}
\graphicspath{{../pdf/}{../jpeg/}}
\DeclareGraphicsExtensions{.pdf,.jpeg,.png}

\usepackage{array}
\usepackage{mdwmath}
\usepackage{mdwtab}
\usepackage{eqparbox}
\usepackage{url}
\usepackage{flushend,cuted}
\usepackage[noend]{algpseudocode}
\usepackage{algorithmicx,algorithm}
\usepackage{subfigure}
\usepackage{caption}
\usepackage{cite}
\usepackage{xcolor}
\usepackage{epstopdf}
\usepackage{cases}
\usepackage{amscd,textcomp,amssymb,epsfig,graphics,epsf,color,amsmath,balance,cite}
\usepackage{multirow}
\usepackage{booktabs}
\usepackage{stfloats}
\usepackage{setspace}
\usepackage{graphicx}
\usepackage{amsmath, amsthm, amsfonts, amssymb, amsbsy}
\usepackage{dsfont}
\usepackage{color}
\usepackage{lipsum} 
\usepackage{bm}
\usepackage{tabu}                     
\usepackage{multirow}                 
\usepackage{multicol}                 
\usepackage{multirow}                
\usepackage{float}                    
\usepackage{makecell}                 
\usepackage{booktabs}                 

\DeclareMathOperator{\vvec}{vec}
\DeclareMathOperator{\tr}{tr}

\newtheorem{lem}{Lemma}

\newtheorem{remark}{Remark}

\allowdisplaybreaks[4]

\begin{document}

\title{\vspace{-0.7cm}\huge{{MMSE Design of RIS-aided Communications}}}

\author{Wen-Xuan Long, \emph{Graduate Student Member, IEEE}, Marco Moretti, \emph{Member, IEEE},\\ Andrea Abrardo, \emph{Senior Member, IEEE}, Luca Sanguinetti, \emph{Senior Member, IEEE}, and Rui Chen, \emph{Member, IEEE}

\thanks{W.-X. Long is with the University of Pisa, Dipartimento di Ingegneria dell'Informazione, Italy, and also with the State Key Laboratory of Integrated Service Networks (ISN), Xidian University, Shaanxi 710071, China (e-mail: wxlong@stu.xidian.edu.cn).}

\thanks{M.~Moretti and L.~Sanguinetti are with the Dipartimento di Ingegneria dell'Informazione, University of Pisa, Pisa, Italy (email: marco.moretti@unipi.it, luca.sanguinetti@unipi.it) and also with CNIT (National Interuniversity Consortium for Telecommunications). Their work here is partially supported by the Italian Ministry of Education and Research (MUR) in the framework of the FoReLab project (Departments of Excellence).}

\thanks{A. Abrardo (abrardo@unisi.it) is with the Department of Information Engineering and Mathematical Sciences of the University of Siena, and also with CNIT. A. Abrardo is supported by the European community within the 6G SHINE project (6G SHort range extreme communication IN Entities).}

\thanks{R. Chen is with the State Key Laboratory of Integrated Service Networks (ISN), Xidian University, Shaanxi 710071, China (e-mail: rchen@xidian.edu.cn). W.-X. Long and R. Chen are partially supported by the National Natural Science Foundation of China under Grant 62271376.}
}
\maketitle

\begin{abstract}
Consider a communication system in which a single-antenna user equipment exchanges information with a multi-antenna base station via a reconfigurable intelligent surface (RIS) in the presence of spatially correlated channels and electromagnetic interference (EMI). To exploit the attractive advantages of RIS technology, accurate configuration of its reflecting elements is crucial. In this paper, we use statistical knowledge of channels and EMI to optimize the RIS elements for {\emph{i})} accurate channel estimation and {\emph{ii})} reliable data transmission. In both cases, our goal is to determine the RIS coefficients that minimize the mean square error, resulting in the formulation of two non-convex problems that share the same structure. To solve these two problems, we present an alternating optimization approach that reliably converges to a locally optimal solution. The incorporation of the diagonally scaled steepest descent algorithm, derived from Newton's method, ensures fast convergence with manageable complexity. Numerical results demonstrate the effectiveness of the proposed method under various propagation conditions. Notably, it shows significant advantages over existing alternatives that depend on a sub-optimal configuration of the RIS and are derived on the basis of different criteria.
\end{abstract}


\vspace{-0.2cm}
\section{Introduction}
Reconfigurable intelligent surfaces (RIS) have received remarkable interest in the context of next-generation wireless systems~\cite{Akyildiz2018,Zhang2019_TWC,DiRenzoJSAC_2020,bjornson2022reconfigurable}. A RIS comprises a planar array of $M$ reflective elements positioned at sub-wavelength intervals. Each element's impedance can be adjusted to introduce a controllable phase-shift to the incoming wave before reflecting it. By optimizing the phase-shift pattern throughout the RIS, it becomes possible to control the reflected wavefront and shape it  into a directed beam aimed at the intended receiver~\cite{bjornson2022reconfigurable}. In a communication system in which a single-antenna user equipment (UE) exchanges information with a multi-antenna base station (BS) via a RIS, accurate estimation of the cascaded channel (from the UE to RIS and from the RIS to the BS) is crucial to properly design the phase-shifts and harness the attractive advantages of the RIS technology. Nevertheless, this is a complicated task, primarily due to passive nature of the RIS and the challenges posed by high-dimensional channels~\cite{Demir2022is,Demir2022Exploiting}. The existence of electromagnetic interference (EMI)~\cite{Electromagnetic2022de,BasarCommLett_2023,Wen-Xuan-CommLett_2023}, which can occur naturally in any environment~\cite{Loyka_2004}, further complicates the task. The aim of this paper is to design the elements of the RIS with the primary goal of initially achieving a precise estimate of the cascaded channel and subsequently enhancing the spectral efficiency (SE) of the system. In contrast to most of previous studies, we consider the existence of EMI.

\subsection{Relevant literature}

The quality of the channel state information (CSI) plays a critical role in determining the performance of RIS-based communications. As a passive device, a RIS cannot estimate the channel locally or actively transmit pilot signals. Therefore, channel estimation must be performed at the BS or UE. In single-user systems, the least-square (LS) estimator \cite{Zheng2020Intelligent,Li2023Channel}, the reduced-subspace LS (RS-LS) estimator \cite{Demir2022Exploiting}, the bilinear generalized approximate message passing algorithm \cite{He2020Cascaded} and the bilinear adaptive vector approximate message passing algorithm \cite{Mirza2021Channel} can be employed to accurately estimate the cascaded channel. In multi-user scenarios, the PARAllel FACtor-based channel estimation framework is introduced in \cite{Wei2021Channel} to efficiently estimate the cascaded channels; while in \cite{Gui2022Channel}, exploiting the linear correlation among multi-user cascaded channels, the estimation of the angle information of the BS-RIS channel is combined with the LS estimator used to estimate the channel gains,  significantly reducing the pilot overhead. Moreover, a two-step multi-user joint channel estimation method based on compressed sensing is proposed in \cite{Chen2023Channel} to acquire CSIs of multiple users simultaneously, taking advantage of  the sparsity of cascaded channels. 

After accurate channel estimation, appropriate RIS phase configuration is also crucial for RIS-aided data transmission. Under the assumption of known CSI, in RIS-aided single-user systems, \cite{Wu2020Beamforming} applies the branch-and-bound method to design the optimal RIS configuration with the criterion of minimizing the transmit power; while \cite{Feng2020Deep} proposes the deep deterministic policy gradient algorithm to obtain the optimal RIS phase shifts to maximize the downlink received SNR. In \cite{Wang2021Joint}, to maximize the ergodic SE, the optimal RIS phase shifts are derived using the semidefinite relaxation algorithm under the premise of shaping the transmitted beamforming with the principal eigenvector of the channel. Similarly, for multi-user systems, the genetic algorithm \cite{Zhi2022Power} and the gradient projection method \cite{Subhash2023Optimal} are applied to design the optimal RIS phase shifts for maximizing the minimum signal-to-interference-plus-noise ratio (SINR), thus ensuring balanced performance among different users. In \cite{Gao2023Robust}, the transmit beamforming vector and RIS phase shifts are alternately optimized using the Lagrangian multiplier method and the distributed alternating direction method of multipliers, respectively, to maximize the sum rate.


As a passive device, the RIS does not amplify signals, but relies on directional reflection. Therefore, to ensure a satisfactory signal-to-noise ratio (SNR) at the receiver, RIS is required to be equipped with a significant surface area for wireless communications. As the size of the RIS increases, it becomes more susceptible to interference \cite{Electromagnetic2022de}. In the presence of constant interference, \cite{Wen-Xuan-CommLett_2023} applies the RS-LS estimator to estimate and remove the interference during cascaded channel estimation. Furthermore, to ensure an acceptable SNR at the receiver, \cite{Khaleel2023Electromagnetic} discusses an EMI elimination scheme for RIS-assisted data transmission, in which the optimal RIS phase shift is obtained by the fitting tool in MATLAB.

\begin{figure}[t]
\setlength{\abovecaptionskip}{-0.1cm}
\setlength{\belowcaptionskip}{-0.3cm}   \begin{center}
\includegraphics[scale=0.9]{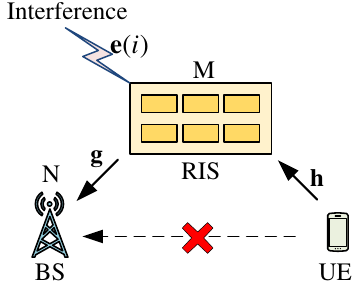}
\end{center}
\caption{RIS-aided communication system.}
\label{Fig1}
\end{figure}

\subsection{Contribution}
As discussed above, reliable RIS-based data transmission depends on accurate CSI. However, existing research on channel estimation and data transmission in RIS-aided systems is generally considered as two separate components, each designed based on different criteria. To the best of our knowledge, there are few researches that jointly consider the channel estimation and data transmission in RIS-aided systems and optimize the overall configuration of RIS based on a unified criteria. Among them, there are \cite{Abrardo-TC-2021} and \cite{Abrardo-TWC-2023}, in which the optimization of the RIS for data transmission is performed based on the statistical knowledge of the node location. The validity of such approaches is therefore limited to cases where the channel is closely related to the node positions, namely in scenarios where there is a strong line of sight (LoS) component in both the BS-RIS and RIS-UE links. In this paper, besides considering more general channel models, we also take into account the presence of EMI, which cannot be generally neglected due to the nature of RIS. In this respect, existing research assumes that the interference experienced by RIS-based communications is approximately constant throughout the entire period, while how to efficiently configure the RIS phase shifts in the presence of random interference for accurate channel estimation and data transmission remains an open issue. To address these issues, this paper aims to propose a joint design framework for RIS-aided communications in the presence of random interference.
In summary, our main contributions are as follows:  
\begin{itemize}
    \item We first derive the linear minimum mean square error (LMMSE) estimator for the cascaded channel and then employ this estimate to formulate the  MMSE combiner at the BS with the aim of improving the SE. In both instances, the efficacy of the estimator and combiner hinges on the phase-shift coefficients of the RIS. To optimize overall performance, we investigate a scheme designed to determine the RIS elements, aiming to minimize the MSE.
    
     \item Since both instances involve solving a non-convex problem to compute the optimal RIS configuration, we provide an iterative algorithm which ensures convergence to a locally optimal solution.
   
     \item The optimization algorithms, employing the principle of \emph{alternating optimization} (AO), leverage the knowledge of the second-order statistics of channels and EMI to implement a projected gradient scheme based on a variant of Newton's method, which balances convergence speed with low computational load.
     
    \item Numerical results validate the method's efficacy across various propagation conditions, highlighting its substantial advantages over existing alternatives relying on suboptimal RIS configurations.
\end{itemize}
\subsection{Paper Outline and Notation}
The remainder of this paper is organized as follows. In Section~\ref{sec:system_model}, we introduce the model adopted for the system, propagation channels and EMI. In Section~III, the channel estimation problem is addressed by means of the LMMSE criterion, followed by the SE computation with different combining schemes including the one minimizing the MSE between the data signal and the received signal
after combining. In Section~IV, we provide an iterative algorithm with provable convergence that allows to optimize the RIS according to the
MSE criteria adopted in the channel estimation and data transmission phases. Numerical results are given in  Section~V while conclusions are drawn in Section VI. 


{\sl Notation}: Unless otherwise specified, matrices are denoted by bold uppercase letters (i.e., $\mathbf{X}$), vectors are represented by bold lowercase letters (i.e., $\mathbf{x}$), and scalars are denoted by normal font (i.e., $x$). $(\cdot)^{\mathrm{T}}$, $(\cdot)^{\mathrm{H}}$ and $(\cdot)^{-1}$ stand for the transpose, Hermitian transpose and inverse of the matrices. The symbol $\otimes$ represents the Kronecker matrix product, while $\odot$ represents the Hadamard product. The notation $\mathbf{x}=\vvec(\mathbf{X})$ defines the linear transformation that converts an $P\times Q$-dimensional $\mathbf{X}$ into a column vector $\mathbf{x}$ with size $PQ\times 1$ by stacking the columns of $\mathbf{X}$ on top of each other, $||\mathbf{x}||$ signifies the Euclidean norm of the vector $\mathbf{x}$, and $\tr\left\{\mathbf{X}\right\}$ indicates the trace of the square matrix $\mathbf{X}$. The notation $\mathbb{E}\{\cdot\}$ represents the statistical expectation, $\mod{(\cdot,\cdot)}$ denotes the modulus operation and $ \lfloor\cdot\rfloor$ truncates the argument.

\section{System Model}\label{sec:system_model}

We consider the RIS-aided communication from  a single-antenna UE to an $N$-element BS, as shown in Fig.\ref{Fig1}. The BS antennas are deployed as a uniform planar array (UPA) with $N_H$ rows and $N_V$ columns,  so that $N = N_HN_V$.  The horizontal and vertical antenna spacing is set to $d_N$.
The RIS is equipped with $M$ passive reconfigurable elements, forming a UPA with $M_H$ rows and $M_V$ columns, where $M=M_HM_V$. The horizontal and vertical element spacing is set to  $d_M$, and the elements are labeled row-by-row as $m\in [1,M]$. Then, the location of the $m$-th element relative to the origin can be expressed as $\textbf{u}_m = [0, \mod(m-1, M_H)d_M, \lfloor(m-1)/M_H\rfloor d_M]^{\text{T}}$ \cite{Bjornson2021Rayleigh}. When a plane-wave impinges on the RIS from the azimuth angle $\varphi$ and elevation angle $\vartheta$, the array response vector can be written as \cite{Demir2022Channel} 
\begin{align}\label{array_vector_a}
\mathbf{a}(\varphi,\vartheta)=\left[e^{j\textbf{k}^{\text{T}}(\varphi,\vartheta)\textbf{u}_1},\ldots,e^{j\textbf{k}^{\text{T}}(\varphi,\vartheta)\textbf{u}_M}\right]^{\text{T}},
\end{align}
where 
\begin{align}
\textbf{k}(\varphi,\vartheta) = \frac{2\pi}{\lambda}[\cos\vartheta\cos\varphi,\cos\vartheta\sin\varphi,\sin\vartheta]^{\text{T}} 
\end{align}
is the wave vector, and $\lambda$ is the wavelength.

\begin{figure}[t]
\setlength{\abovecaptionskip}{-0.1cm}
\setlength{\belowcaptionskip}{-0.3cm}   \begin{center}
\includegraphics[scale=0.90]{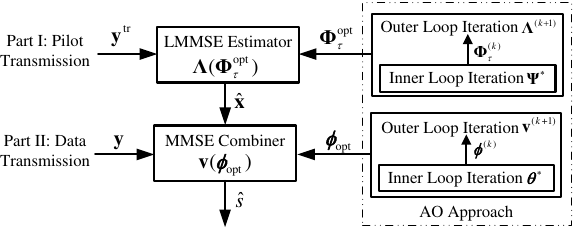}
\end{center}
\caption{Block diagram of MMSE receiver in the RIS-aided communication. The receiver operation is divided in two parts: (I) channel estimation, and (II) data
transmission.}
\label{Fig2}
\end{figure}

We consider a block-fading model where each channel takes one realization in a coherence block of $\tau_c$ channel uses and independent realizations across blocks.
The channel from the UE to the RIS is called $\mathbf{h}$ and modelled as $\mathbf{h}\sim\mathcal{N}_{\mathbb{C}}(\mathbf{0}_M,\mathbf{R}_h)$, 
with $\mathbf{R}_h\in\mathbb{C}^{M\times M}$ being the spatial correlation matrix \cite{Bjornson2021Rayleigh,Demir2022Channel}:
\begin{align}\label{Rh}
\mathbf{R}_h=\beta_h\!\int\!\!\!\!\int^{\pi/2}_{-\pi/2}f_h(\varphi,\vartheta)\mathbf{a}(\varphi,\vartheta)\mathbf{a}^{\text{H}}(\varphi,\vartheta)d\vartheta d\varphi,
\end{align}
where $\beta_h$ is the channel gain, and $f_h(\varphi,\vartheta)$ is the normalized spatial scattering function with $\int\!\!\!\int f_h(\varphi,\vartheta)d\vartheta d\varphi=1$.

The narrowband channel from the $m$-th RIS element to the $n$-th BS element is indicated as $g_{m,n}$. We call $\mathbf{g}_n=[g_{n,1},g_{n,2},\ldots,g_{n,M}]^{\text{T}}\in \mathbb{C}^M$ the channel vector from the RIS to the $n$-th BS element, while $\mathbf{g}'_m$ $=$ $[g_{1,m},g_{2,m},\cdots,g_{N,m}]^{\text{T}}\in \mathbb{C}^N$  denotes the channel vector from the $m$-th RIS element to the BS array. Using the Kronecker model \cite{Demir2022is,DaShan2000Fading}, we have that 
\begin{align}
\mathbf{g}_n&\sim \mathcal{N}_{\mathbb{C}}(\mathbf{0}_M,[\mathbf{R}_{g'_m}]_{n,n}\mathbf{R}_{g_n})\\
\mathbf{g}'_m&\sim \mathcal{N}_{\mathbb{C}}(\mathbf{0}_N,[\mathbf{R}_{g_n}]_{m,m}\mathbf{R}_{g'_m})
\end{align}
where $[\mathbf{R}_{g'_m}]_{n,n}\mathbf{R}_{g_n}$ and $[\mathbf{R}_{g_n}]_{m,m}\mathbf{R}_{g'_m}$ are the spatial correlation matrices of $\mathbf{g}_n$ and $\mathbf{g}'_m$, respectively. Both $\mathbf{R}_{g_n}$ and $\mathbf{R}_{g'_m}$ have the same form as \eqref{Rh} but must be computed using the specific spatial scattering function and channel gain corresponding to $\mathbf{g}_n$ and $\mathbf{g}'_m$.

We assume the channels $\mathbf{g}_n$ (and thus $\mathbf{g}'_m$) and $\mathbf{h}$ are independent of each other.
The cascaded channel between the UE and the $n$-th element of the BS can be written as an $M$-dimensional vector
\begin{equation}
\mathbf{x}_{n}=\mathbf{g}_{n}\odot\mathbf{h}.
\end{equation}

A common practice in existing RIS-aided communications is to only consider the signals generated by the system, and thereby neglecting the EMI or “noise” (or “pollution”) that is inevitably present in any environment. The EMI may arise from a variety of natural, intentional or non-intentional causes, for example, man-made devices and natural background radiation.
In this paper, we call $\mathbf{e}(i)\in \mathbb{C}^M$ the vector collecting the EMI during the $i$-th channel use, which accounts for any uncontrollable factor (e.g., of electromagnetic or hardware nature) disturbing the incoming signals at the RIS. We model it as $\mathbf{e}(i)\sim \mathcal{N}_{\mathbb{C}}(\mathbf{0}_M,\sigma_{e}^{2}\mathbf{R}_e)$ and assume that it takes independent realizations across channel uses, i.e., $\mathbb{E}\{\mathbf{e}(i)\mathbf{e}(i')^{\text{H}}\}=0$ for $i\neq i'$. The normalized spatial correlation matrix $\mathbf{R}_e\in\mathbb{C}^{M\times M}$ has the same form of \eqref{Rh}, i.e.,
\begin{align}\label{Rn}
\mathbf{R}_e=\int\!\!\!\!\int^{\pi/2}_{-\pi/2}f_e(\varphi,\vartheta)\mathbf{a}(\varphi,\vartheta)\mathbf{a}^{\text{H}}(\varphi,\vartheta)d\vartheta d\varphi,
\end{align}
but with a different spatial scattering function $f_e(\varphi,\vartheta)$.

\section{System Operation}
\vspace{-0.0cm}

We assume that the system operates according to a communication protocol wherein the data transmission phase is preceded by a training phase for channel estimation, as shown in Fig. \ref{Fig2}, where the $\tau_c$ available channel uses are employed for: $\tau$ for training phase and $\tau_c - \tau$ for uplink payload transmission. We assume that no direct link is present between UE and BS.

\subsection{Pilot Transmission and Linear MMSE Estimation}

We assume that the training sequence of length $\tau$ is composed by all ones and denote $\bm{\phi}(i)\in \mathbb{C}^{M}$ the vector collecting the controllable phase-shifts $\{{\phi}_m(i)\in[0,2\pi); m=1,\ldots,M\}$ introduced by the RIS during the $i$-th channel use. The training signal $y_{n}^{\mathrm{tr}}(i)$ received by the $n$-th BS antenna takes the form:
\begin{align}\label{ym2}
y_{n}^{\mathrm{tr}}(i)=\sqrt{\rho^{\mathrm{tr}}}\bm{\phi}(i)^{\mathrm{T}}\mathbf{x}_{n}+w_n (i)+z_n(i),
\end{align}
where $\rho^{\mathrm{tr}}$ is the power of the training signal while
\begin{align}
w_n(i) = \bm{\phi}(i)^{\mathrm{T}}\left(\mathbf{g}_n \odot\mathbf{e}(i)\right),
\end{align}
and $z_n(i)\sim \mathcal{N}_{\mathbb{C}}(0,\sigma^2)$ is the additive white Gaussian noise. Notice that the term $w_n(i)$ depend on the RIS configuration. By collecting all the training signals received at the $n$-th BS antenna during the $\tau$ channel uses of the training phase, we obtain $\mathbf{y}_{n}^{\mathrm{tr}}=[y^{\mathrm{tr}}_n(1),\ldots,y^{\mathrm{tr}}_n({\tau})]^{\mathrm{T}}\in\mathbb{C}^{\tau}$ given by
\begin{align}\label{y2.1}
\mathbf{y}_{n}^{\mathrm{tr}}=\sqrt{\rho^{\mathrm{tr}}}\mathbf{\Phi}_{\tau}\mathbf{x}_{n}+\mathbf{w}_{n}^{\mathrm{tr}}+\mathbf{z}_{n}^{\mathrm{tr}},
\end{align}
where
\begin{align} 
 \mathbf{\Phi}_{\tau} = [\bm{\phi}(1),\bm{\phi}(2),\cdots,\bm{\phi}(\tau)]^{\mathrm{T}}\in \mathbb{C}^{\tau \times M},
 \end{align}
$\mathbf{w}_{n}^{\mathrm{tr}}=[w_n(1),\ldots,w_n({\tau})]^{\mathrm{T}}$ and $\mathbf{z}_{n}^{\mathrm{tr}}=[z_n(1),\ldots,z_n({\tau})]^{\mathrm{T}}$. 
 Accordingly, the  vector $\mathbf{y}^{\mathrm{tr}}= [{\mathbf{y}^{\mathrm{tr}}_{1}}^{\mathrm{T}}, \ldots, {\mathbf{y}^{\mathrm{tr}}_{N}}^{\mathrm{T}}]^{\mathrm{T}}$ $\in\mathbb{C}^{N\tau }$, obtained by collecting the signals received at the BS array during the training phase, takes the form
\begin{align}
\label{y2.2}
\mathbf{y}^{\mathrm{tr}}=\sqrt{\rho^{\mathrm{tr}}}\mathbf{\Phi}_{N\tau}\mathbf{x}+\mathbf{w}^{\mathrm{tr}}+\mathbf{z}^{\mathrm{tr}}
\end{align}
with 
\begin{align} 
\mathbf{\Phi}_{N{\tau}} = \mathbf{I}_N\otimes\mathbf{\Phi}_{\tau},
\end{align}
and $\mathbf{x}= [\mathbf{x}^{\mathrm{T}}_{1}, \ldots, \mathbf{x}^{\mathrm{T}}_{N}]^{\mathrm{T}}\in\mathbb{C}^{MN }$, $\mathbf{w}^{\mathrm{tr}}$ $=$ $ [{\mathbf{w}^{\mathrm{tr}}_{1}}^{\mathrm{T}},$$ \ldots,$ $ {\mathbf{w}^{\mathrm{tr}}_{N}}^{\mathrm{T}}]^{\mathrm{T}}\in\mathbb{C}^{N\tau }$ and $\mathbf{z}^{\mathrm{tr}}= [{\mathbf{z}^{\mathrm{tr}}_{1}}^{\mathrm{T}}, \ldots, {\mathbf{z}^{\mathrm{tr}}_{N}}^{\mathrm{T}}]^{\mathrm{T}}\in\mathbb{C}^{N\tau }$.

We assume that the BS has knowledge of the correlation matrices $\mathbf{R}_{x}= \mathbb{E}\{\mathbf{x}\mathbf{x}^{\mathrm{H}}\}$ and $\mathbf{R}_{w}^{\mathrm{tr}} = \frac{1}{\sigma^{2}_{e}}\mathbb{E}\{\mathbf{w}^{\mathrm{tr}}{(\mathbf{w}^{\mathrm{tr}})}^{\mathrm{H}}\}$. These are given by (see Appendix A)
\begin{align}\label{Rw_prime}
\mathbf{R}_{x}=\mathbf{R}_{g'_m}\otimes\left(\mathbf{R}_{g_n}\odot\mathbf{R}_{h}\right)=\mathbf{R}_{g'_m}\otimes\mathbf{R}_{c}
\end{align}
and
\begin{align}\label{Rw_prime}
\mathbf{R}_{w}^{\mathrm{tr}} =  \mathbf{R}_{g'_m}\otimes\left(\!\left(\mathbf{\Phi}_{{\tau}}\mathbf{R}_{q}\mathbf{\Phi}_{{\tau}}^{\mathrm{H}}\right)\odot\mathbf{I}_{\tau}\right), 
\end{align}
where we have defined
\begin{align}\label{R_c}
\mathbf{R}_{c}=\mathbf{R}_{g_n}\odot\mathbf{R}_{h}\in\mathbb{C}^{M\times M}
\end{align}
and 
\begin{align}\label{R_q}
\mathbf{R}_{q} = \mathbf{R}_{g_n}\odot\mathbf{R}_{e}\in\mathbb{C}^{M\times M}
\end{align}
for subsequent use. If $\tau =1$, then \eqref{Rw_prime} reduces to 
\begin{align}\label{Rw_prime_1}
\mathbf{R}_{w}^{\mathrm{tr}} =  \mathbf{R}_{g'_m}\otimes\!\left(\bm{\phi}^{\mathrm{T}}\mathbf{R}_{q}\bm{\phi}^{*}\right),
\end{align}
from which, by applying $(\mathbf{AC}) \otimes (\mathbf{BD})=(\mathbf{A} \otimes \mathbf{B})(\mathbf{C} \otimes \mathbf{D})$ twice, we get
\begin{align}\label{Rw_prime_2}
\mathbf{R}_{w}^{\mathrm{tr}}= \mathbf{\Phi}_{N} \left(\mathbf{R}_{g'_m}\otimes\mathbf{R}_{q}\right)\mathbf{\Phi}_{N}^{\mathrm{H}}
\end{align}
with
\begin{align}
\mathbf{\Phi}_{N} = \mathbf{I}_N\otimes\bm{\phi}^{\mathrm{T}}.
\end{align}
The statistics above are used to compute $\hat{\mathbf{{x}}}$, the LMMSE estimate of $\mathbf{{x}}$ based on $\mathbf{y}^{\mathrm{tr}}$. 

\begin{lem}
The LMMSE estimate of $\mathbf{x}$ based on the observation of $\mathbf{y}^{\mathrm{tr}}$ is
    \begin{align}\label{LMMSE_estimator}
    \hat{\mathbf{{x}}} = \frac{1}{\sqrt{\rho^{\mathrm{tr}}}}\mathbf{R}_{x}\mathbf{\Phi}_{N{\tau}}^{\mathrm{H}}\left({{\bf R}_y^{\mathrm{tr}}}\right)^{-1}\mathbf{y}^{\mathrm{tr}},
    \end{align}
    where ${{\bf R}_y^{\mathrm{tr}}}= \frac{1}{{\rho^{\mathrm{tr}}}}\mathbb{E}\{\mathbf{y}^{\mathrm{tr}}{(\mathbf{y}^{\mathrm{tr}})}^{\mathrm{H}}\} $ is given by
\begin{align}\label{R_y_tr}
{{\bf R}_y^{\mathrm{tr}}}=\mathbf{\Phi}_{N{\tau}}\mathbf{R}_{x}\mathbf{\Phi}_{N{\tau}}^{\mathrm{H}}+\frac{\sigma^{2}_{e}}{\rho^{\mathrm{tr}}}\mathbf{R}_{w}^{\mathrm{tr}}+\frac{\sigma^2}{\rho^{\mathrm{tr}}}\mathbf{I}_{N\tau}.
\end{align}
The estimation error $\tilde{\mathbf{{x}}} = {\mathbf{{x}}}  - \hat{\mathbf{{x}}} $ has correlation matrix
\begin{equation}\label{corr_R_tildex}
\mathbf{R}_{\tilde x}= \mathbb{E}\{\tilde{\mathbf{{x}}}\tilde{\mathbf{{x}}}^{\mathrm{H}}\} = \mathbf{R}_{x}-\mathbf{R}_x\mathbf{Q}^{\mathrm{tr}}(\mathbf{\Phi}_{{\tau}})\mathbf{R}_x
\end{equation}
with
\begin{equation}
\mathbf{Q}^{\mathrm{tr}}(\mathbf{\Phi}_{{\tau}}) = \mathbf{\Phi}_{N{\tau}}^{\mathrm{H}}\left({{\bf R}_y^{\mathrm{tr}}}\right)^{-1}\mathbf{\Phi}_{N \tau}.
\end{equation}
\end{lem}

\begin{IEEEproof}
The proof follows from standard arguments and is given in Appendix A for completeness.
\end{IEEEproof}

\smallskip
From \eqref{corr_R_tildex}, the MSE is given by
\begin{equation}\label{MSE_MMSE}
\mathcal{E}_x(\mathbf{\Phi}_{{\tau}}) = \text{tr}\left\{\mathbf{R}_{\tilde x}\right\} = \text{tr}\left\{\mathbf{R}_{x}-\mathbf{R}_x\mathbf{Q}^{\mathrm{tr}}(\mathbf{\Phi}_{{\tau}})\mathbf{R}_x\right\},
\end{equation}
which depends only on ${\mathbf{\Phi}_{{\tau}}}$ since the channel statistics are fixed and given.

\begin{remark}
A scheme that dispenses from knowledge of statistics is the LS channel estimator {\cite{Demir2022Channel}}. In this case, we have that
    \begin{align}\label{LS_estimator}
    \hat{\mathbf{{x}}} = \frac{1}{\sqrt{\rho^{\mathrm{tr}}}}\mathbf{A}_1\mathbf{y}^{\mathrm{tr}}
    \end{align}
    with $\mathbf{A}_1 = \left(\mathbf{\Phi}_{N\tau}^{\mathrm{H}}\mathbf{\Phi}_{N\tau}\right)^{-1}\mathbf{\Phi}_{N\tau}^{\mathrm{H}}$, and the MSE takes the form
    \begin{align}\label{MSE_LS}
     \mathcal{E}'_x(\mathbf{\Phi}_{{\tau}}) = \rm{tr}\left\{{\mathbf{R}}_{x}-2\Re{\rm{e}}\left({\bf A}_1\mathbf{\Phi}_{N{\tau}}{\mathbf{R}}_x\right)+ {\bf A}_1{{\bf R}_y^{\mathrm{tr}}}{\bf A}_1^{\mathrm{H}}\right\}.
\end{align}
By using \eqref{R_y_tr} and ${\bf A}_1\mathbf{\Phi}_{N{\tau}} = {\bf I}_{N \tau}$, we obtain
    \begin{equation}\label{MSE_LS}
    {{\mathcal{E}'_x(\mathbf{\Phi}_{{\tau}}) =\mathrm{tr}\left\{\frac{\sigma^{2}_{e}}{\rho^{\mathrm{tr}}}{\mathbf{A}}_1\mathbf{R}_{w}^{\mathrm{tr}}{\mathbf{A}}_1^{\mathrm{H}}+\frac{\sigma^2}{\rho^{\mathrm{tr}}}\left(\mathbf{\Phi}_{N\tau}^{\mathrm{H}}\mathbf{\Phi}_{N\tau}\right)^{-1}\right\}}.}
\end{equation}
Alternatively, one may resort to the RS-LS estimate, which can be obtained by employing the reduced-subspace linear filter \cite{Demir2022Exploiting}, i.e.,
\begin{align}\label{RSLS_estimator}
    \hat{\mathbf{{x}}} = \frac{1}{\sqrt{\rho^{\mathrm{tr}}}}\mathbf{A}_2\mathbf{y}^{\mathrm{tr}}
    \end{align} 
where     
\begin{align}\label{RSLS_A2}
    \mathbf{A}_2 = \mathbf{U}_s \left(\mathbf{U}_s^{\mathrm{H}}\mathbf{\Phi}_{N\tau}^{\mathrm{H}}\mathbf{\Phi}_{N\tau}\mathbf{U}_s\right)^{-1}\mathbf{U}_s^{\mathrm{H}}\mathbf{\Phi}_{N\tau}^{\mathrm{H}},
\end{align}
 and $\mathbf{U}_s\in\mathbb{C}^{MN\times r}$ spans the signal subspace of $\mathbf{R}_x$ containing $\mathbf{{x}}$, and $r = \text{rank}\{\mathbf{R}_x\}$. The MSE is given by
\begin{equation}\label{MSE_RSLS}
    {{\mathcal{E}''_x(\mathbf{\Phi}_{{\tau}}) =\mathrm{tr}\left\{\frac{\sigma^{2}_{e}}{\rho^{\mathrm{tr}}}{\mathbf{A}}_2\mathbf{R}_{w}^{\mathrm{tr}}{\mathbf{A}}_2^{\mathrm{H}}+\frac{\sigma^2}{\rho^{\mathrm{tr}}}\left(\mathbf{U}_s^{\mathrm{H}}\mathbf{\Phi}_{N\tau}^{\mathrm{H}}\mathbf{\Phi}_{N\tau}\mathbf{U}_s\right)^{-1}\right\}}.}
\end{equation}
When $r< MN$, the RS-LS estimator offers superior performance compared to the LS estimator due to noise removal from the unused channel dimensions \cite{Demir2022Exploiting}.
\end{remark}

\begin{remark}
The LMMSE estimator in \eqref{LMMSE_estimator} can be applied for any $\tau \ge 1$, but better results are achieved for larger values of $\tau$. The LS estimator in \eqref{LS_estimator} requires $\tau\geq M$, which may be large, whereas the RS-LS estimator in \eqref{RSLS_estimator} relaxes the required number of pilots to $\tau \geq \frac{r}{N}$ \cite{Demir2022Exploiting}.
\end{remark}

\subsection{Spectral Efficiency and MMSE Combining}
Similar to \eqref{ym2}, the signal received at the $n$-th antenna during data transmission is
\begin{align}\label{ym2_data}
y_{n}=\sqrt{\rho}\bm{\phi}^{\mathrm{T}}\mathbf{x}_{n} s+w_n+z_n,
\end{align}
where $\rho$ is the power of the transmitted signal $s \in \mathbb{C}$ and
\begin{align}
w_n = \bm{\phi}^{\mathrm{T}}\left(\mathbf{g}_n \odot\mathbf{e}\right).
\end{align}
The received vector $\mathbf{y} = [y_1,\ldots,y_N]^{\mathrm{T}} \in \mathbb{C}^{N}$ at the BS takes the form
\begin{align}
\label{y2.2}
\mathbf{y}=\sqrt{\rho}\mathbf{\Phi}_N\mathbf{x} s +\mathbf{w}+\mathbf{z},
\end{align}
where $\mathbf{z} \sim \mathcal{N}_{\mathbb{C}}(\mathbf{0},\sigma^2 \mathbf{I}_N)$ while $\mathbf{w}=[w_1,\ldots,w_N]^{\mathrm{T}}$ is such that
\begin{align}
{\bf R}_w= \frac{1}{\sigma^{2}_{e}}\mathbb{E}\{\mathbf{w}\mathbf{w}^{\mathrm{H}}\} = \mathbf{\Phi}_{N} \left(\mathbf{R}_{g'_m}\otimes\mathbf{R}_{q}\right)\mathbf{\Phi}_{N}^{\mathrm{H}}
\end{align}
with ${\bf R}_q$ being given by \eqref{R_q}.
The BS estimates the signal $s$ by using the combining vector $\bf v$ to obtain 
\begin{align}\label{v_1.0}
\hat{s}={\bf v}^{\mathrm{H}}\mathbf{y} = \sqrt{\rho}{\bf v}^{\mathrm{H}}\mathbf{\Phi}_N\mathbf{x} s +{\bf v}^{\mathrm{H}}\mathbf{w}+{\bf v}^{\mathrm{H}}\mathbf{z}.
\end{align}
In the ideal case of perfect knowledge of the channel, i.e., $\hat{\mathbf{x}}=\mathbf{x}$, the SE of the system is computed as (\!\!\cite[Th. 4.1]{Bjornson2017Massive}) 
\begin{align}\label{SE_1}
\mathrm{SE} & = \frac{\tau_c-\tau}{\tau_c} \mathbb{E}\left\{\log_2\left(1 + \mathrm{SINR}\right)\right\},
\end{align}
with
\begin{align}\label{SINR}
\mathrm{SINR} = \frac{\left|{\bf v}^{\mathrm{H}}\mathbf{\Phi}_N\mathbf{x}\right|^2}{{\bf v}^{\mathrm{H}}\left(\frac{\sigma_e^2}{\rho}{\mathbf{R}_w} +\frac{\sigma^2}{\rho}\mathbf{I}_N\right){\bf v}},
\end{align}
where the expectation is computed with respect to the channel realizations and the pre-log factor  accounts for the fraction of data channel uses per coherence block. The above expression holds for any choice of the combiner. 
From \eqref{SINR}, we see that the maximization of the SINR is a generalized Rayleigh quotient. Thus, the optimal combiner is (\!\!~\cite[App. C.3.2]{Bjornson2017Massive}) 
\begin{align}\label{MR_withEMI}
\mathbf{v} = \mu\left(\frac{\sigma_e^2}{\rho}{\mathbf{R}_w} +\frac{\sigma^2}{\rho}\mathbf{I}_N\right)^{-1}\!\!\!\!\mathbf{\Phi}_N{\mathbf{{x}}},
\end{align}
where $\mu$ is an arbitrary non-zero scalar factor that does not change the SINR in \eqref{SINR}. Plugging \eqref{MR_withEMI} into \eqref{SINR} yields
\begin{align}\label{SINR_2}
\mathrm{SINR} =  \mathbf{{x}}^{\mathrm{H}}\mathbf{\Phi}_N^{\mathrm{H}}\left(\frac{\sigma_e^2}{\rho}{\mathbf{R}_w} +\frac{\sigma^2}{\rho}\mathbf{I}_N\right)^{-1}\!\!\!\mathbf{\Phi}_N{\mathbf{{x}}}.
\end{align}
By choosing 
\begin{align}
\mu = \frac{1}{\sqrt{\rho}}\frac{1}{1+\mathbf{{x}}^{\mathrm{H}}\mathbf{\Phi}_N^{\mathrm{H}}\left(\frac{\sigma_e^2}{\rho}{\mathbf{R}_w} +\frac{\sigma^2}{\rho}\mathbf{I}_N\right)^{-1}\!\!\!\mathbf{\Phi}_N \mathbf{{x}}},
\end{align}
we obtain\footnote{By using $\frac{1}{1+ \mathbf{a}^{\mathrm{H}}\mathbf{B}^{\mathrm{-1}}\mathbf{a}}\mathbf{B}^{\mathrm{-1}}\mathbf{a} = (\mathbf{B} + \mathbf{a}\mathbf{a}^{\mathrm{H}})^{\mathrm{-1}}\mathbf{a}$ with $\mathbf{B} = \frac{\sigma_e^2}{\rho}{\mathbf{R}_w} +\frac{\sigma^2}{\rho}\mathbf{I}_N$ and $\mathbf{a}=\mathbf{\Phi}_N{\mathbf{{x}}}$.}
\begin{align}\label{MMSE_withEMI}
\mathbf{v} &= \frac{1}{\sqrt{\rho}}\left(\mathbf{\Phi}_N \mathbf{{x}}\mathbf{{x}}^{\mathrm{H}}\mathbf{\Phi}_N^{\mathrm{H}}+\frac{\sigma_e^2}{\rho}{\mathbf{R}_w} +\frac{\sigma^2}{\rho}\mathbf{I}_N\right)^{-1}\!\!\!\!\mathbf{\Phi}_N{\mathbf{{x}}},
\end{align}
which corresponds to the vector minimizing the MSE $\mathcal{E}_{s}= \mathbb{E}\{|\mathbf{v}^{\mathrm{H}} \mathbf{y}\!-\!s |^2\}$ between the data signal and the received signal $\mathbf{v}^{\mathrm{H}} \mathbf{y}$ after combining. Since $\mu$ does not alter the SINR in \eqref{SINR_2}, it thus follows that the two combiners in \eqref{MR_withEMI} and \eqref{MMSE_withEMI} are equivalent in terms of achievable SE. By using standard calculus, \eqref{SINR_2} can be equivalently expressed as
\begin{equation}\label{equivalence_SINR_E}    
\mathrm{SINR} = \frac{1}{\mathcal{E}_{s}}-1.
\end{equation}
%

The characterization of the SE with imperfect channel knowledge requires to follow a different approach. An achievable SE 
can be computed using the so-called hardening capacity bound, which has received great attention in the massive MIMO literature, e.g., \!~\cite[Sec. 4.2]{Bjornson2017Massive}). By adding and subtracting $\sqrt{\rho}\mathbb{E}\left\{{\bf v}^{\mathrm{H}}\mathbf{\Phi}_N\mathbf{x}\right\}s$, the combined signal ${\bf v}^{\mathrm{H}}\mathbf{y}$ can be rewritten as
\begin{align}
{\bf v}^{\mathrm{H}}\mathbf{y} = \sqrt{\rho}\mathbb{E}\left\{{\bf v}^{\mathrm{H}}\mathbf{\Phi}_N\mathbf{x}\right\}s  + {\xi},
\end{align}
where 
\begin{align}
{\xi} = \sqrt{\rho}\left({\bf v}^{\mathrm{H}}\mathbf{\Phi}_N\mathbf{x} -\mathbb{E}\left\{{\bf v}^{\mathrm{H}}\mathbf{\Phi}_N\mathbf{x}\right\}\right)s +{\bf v}^{\mathrm{H}}\mathbf{w}+{\bf v}^{\mathrm{H}}\mathbf{z}
\end{align}
has zero mean, i.e., $\mathbb{E}\left\{{\xi}\right\} = 0$ and is uncorrelated with the input $s$, i.e., $\mathbb{E}\left\{{\xi s^{*}}\right\} = 0$. In this case, an achievable SE can be obtained as (\!\!\cite[App. C.3.4]{Bjornson2017Massive})
\begin{align}\label{SE_2}
\!\!\!\!\mathrm{SE} & = \!\frac{\tau_c-\tau}{\tau_c} \!\log_2\left(1 + \frac{\left|\mathbb{E}\left\{{\bf v}^{\mathrm{H}}\mathbf{\Phi}_N\mathbf{x}\right\}\right|^2}{\mathbb{E}\left\{|{\xi}|^2\right\}}\right)
\end{align}
with
\begin{align}
{\mathbb{E}\left\{|{\xi}|^2\right\}} = &\mathbb{E}\left\{\left|{\bf v}^{\mathrm{H}}\mathbf{\Phi}_N\mathbf{x}\right|^2\right\}-\left|\mathbb{E}\left\{{\bf v}^{\mathrm{H}}\mathbf{\Phi}_N\mathbf{x}\right\}\right|^2 \\&+\frac{\sigma^{2}_{e}}{\rho}\mathbb{E}\left\{{\bf v}^{\mathrm{H}}{\bf R}_w{\bf v}\right\} + \frac{\sigma^{2}}{\rho}\mathbb{E}\left\{||\mathbf{v}||^2\right\}\nonumber
\end{align}
and the expectations are computed with respect to all sources of randomness. The achievable SE above can also be computed with any combining scheme and channel estimator. We use the MMSE combiner that minimizes the MSE given by $\mathcal{E}_{s}= \mathbb{E}\{|\mathbf{v}^{\mathrm{H}} \mathbf{y}\!-\!s |^2\big|\hat{\mathbf{x}}\}$, where $\mathbb{E}\{\cdot\big|\hat{\mathbf{x}}\}$ denotes the  expectation conditioned to the channel estimate $\hat{\mathbf{x}}$. From \eqref{v_1.0}, we obtain
\begin{equation}
\begin{aligned}\label{MSE_s11}
\mathcal{E}_{s} = 1+\rho\mathbf{v}^{\mathrm{H}}\mathbf{R}_y\mathbf{v} -2\sqrt{\rho} \Re{\rm{e}}\left(\mathbf{v}^{\mathrm{H}}\mathbf{\Phi}_N\hat{\mathbf{x}}\right),
\end{aligned}
\end{equation}
with ${\bf R}_y= \frac{1}{\rho}\mathbb{E}\{\mathbf{y}\mathbf{y}^{\mathrm{H}}\} $ given by 
\begin{align}
{\bf R}_y= \boldsymbol{\Phi}_N\left( \hat{\mathbf{x}}\hat{\mathbf{x}}^{\mathrm{H}}+\mathbf{R}_{\tilde x}\right)\boldsymbol{\Phi}_N^{\mathrm{H}} +\frac{\sigma_e^2}{\rho}{\mathbf{R}_w} +\frac{\sigma^2}{\rho}\mathbf{I}_N.
\end{align}
Minimizing \eqref{MSE_s11} with respect to $\mathbf{v}$ yields
\begin{equation} \label{opt_2.120}
\mathbf{v} =\frac{1}{\sqrt{\rho}} \mathbf{R}_y^{-1} \boldsymbol{\Phi}_N \hat{\mathbf{x}},  
\end{equation}
so that 
the minimum MSE is given by 
\begin{equation}
\label{MSE_MMSE_s1_1}
\begin{aligned}
\mathcal{E}_{s}(\bm{\phi}) = 1 -\hat{\mathbf{x}}^{\mathrm{H}}\mathbf{Q}(\bm{\phi})\hat{\mathbf{x}}    
\end{aligned}
\end{equation}
with
\begin{equation}
\mathbf{Q}(\bm{\phi}) = \boldsymbol{\Phi}_N^{\mathrm{H}}\mathbf{R}_y^{-1}\boldsymbol{\Phi}_N.
\end{equation}
Similar to \eqref{MSE_MMSE}, the MSE \eqref{MSE_MMSE_s1_1} depends only on $\bm{\phi}$ and the RIS coefficients can be optimized to minimize it.

\section{MMSE Optimization of the RIS}
We now present an algorithm designed to optimize the RIS according to the MMSE criterion in the channel estimation and data transmission phases. 

\subsection{Optimizing RIS for Channel Estimation}
The channel estimation phase is considered first.
To  minimize the channel estimation error in \eqref{MSE_MMSE}, we can exploit the dependence on $\mathbf{\Phi}_{\tau}$, and design it to solve the following optimization problem:
\begin{equation}
\begin{aligned} \label{opt_1}
&\!\min_{\mathbf{\Phi}_{\tau}\in\mathcal{F}}\ \mathcal{E}_x(\mathbf{\Phi}_{\tau})
\end{aligned}
\end{equation}
where the feasible set 
\begin{equation}
\mathcal{F}=\{\mathbf{\Phi}_{\tau} \in \mathbb{C}^{\tau\times M}\vert \ |[\mathbf{\Phi}_{\tau}]_{i,m}|=1;\forall i,m\}
\end{equation}
captures the fact that the RIS is a passive device whose coefficients must have unitary modulus. 

\begin{algorithm}[t]
\caption{The AO outer loop iteration}
\hspace*{0.02in} {\bf Input:}
Desirable accuracy $\epsilon$ and the initial phase-shifts $\mathbf{\Phi}^{(0)}_{\tau}$\\
\hspace*{0.02in} {\bf Output:}
Optimized RIS phase-shift matrix $\mathbf{\Phi}^{\text{opt}}_{\tau}$
\begin{algorithmic}[1]
\State {\bf Init}. initialize $k = 0$ and $\Delta_{\mathcal{E}} = 1$.
\State $\mathbf{\Lambda}^{(k+1)} \leftarrow$ substitute $\mathbf{\Phi}_{\tau}^{(k)}$ into \eqref{Lamda_Phi};
\State $\mathcal{E}_{x}\left(\mathbf{\Lambda}^{(k+1)},\mathbf{\Phi}_{\tau}\right) \leftarrow$ substitute $\mathbf{\Lambda}^{(k+1)}$ into \eqref{MSE_k_lambda};
\State {\bf while} $\frac{\Delta_{\mathcal{E}}}{\mathcal{E}_{x}\left(\mathbf{\Lambda}^{(k+1)},\mathbf{\Phi}_{\tau}\right)}>\epsilon$ {\bf do}
\State$\ \ $$\Delta_{\mathcal{E}}\leftarrow \mathcal{E}_{x}\left(\mathbf{\Lambda}^{(k+1)},\mathbf{\Phi}_{\tau}\right)$;
\State$\ \ $Algorithm 2 $\leftarrow\mathbf{\Phi}_{\tau}^{(k)}$;
\State$\ \ $$\mathbf{\Psi}^{\star}\leftarrow$ Perform Algorithm 2;
\State$\ \ $$\mathbf{\Phi}_{\tau}^{(k+1)}\leftarrow e^{j\angle{\mathbf{\Psi}^{\star}}}$;
\State$\ \ $$k = k +1$;
\State$\ \ $$\mathbf{\Lambda}^{(k+1)}$ $\leftarrow$ substitute $\mathbf{\Phi}_{\tau}^{(k)}$ into \eqref{Lamda_Phi};
\State$\ \ $$\mathcal{E}_{x}\left(\mathbf{\Lambda}^{(k+1)},\mathbf{\Phi}_{\tau}\right)\leftarrow$  substitute $\mathbf{\Lambda}^{(k+1)}$ into \eqref{MSE_k_lambda};
\State$\ \ $$\Delta_{\mathcal{E}}= \mathcal{E}_{x}\left(\mathbf{\Lambda}^{(k+1)},\mathbf{\Phi}_{\tau}\right)-\Delta_{\mathcal{E}}$;
\State {\bf end while}
\State $\mathbf{\Phi}^{\text{opt}}_{\tau} \leftarrow \mathbf{\Phi}_{\tau}^{(k)}$.
\State \textbf{end procedure}
\end{algorithmic}\label{alg1}
\end{algorithm}

The optmization problem \eqref{opt_1} is not convex in $\mathbf{\Phi}_{\tau}$ and finding its solution involves the optimization over large matrices. To solve it, we rewrite \eqref{MSE_MMSE} as
\begin{equation}\label{MSE_MMSE2}
\mathcal{E}_x(\mathbf{\Phi}_{{\tau}}) = \text{tr}\left\{\mathbf{R}_{x}-\mathbf{\Lambda}(\mathbf{\Phi}_{{\tau}})\mathbf{\Phi}_{{N\tau}}\mathbf{R}_x\right\},
\end{equation}
with
    \begin{align}\label{Lamda_Phi}
    \mathbf{\Lambda}(\mathbf{\Phi}_{\tau}) =\mathbf{R}_{x}\mathbf{\Phi}_{N{\tau}}^{\mathrm{H}}\left({{\bf R}_y^{\mathrm{tr}}}\right)^{-1}.
    \end{align}
Moreover, we make a simplifying assumption, that is, we neglect the fact that $\mathbf{\Lambda}(\mathbf{\Phi}_{\tau})$ depends on the value of $\mathbf{\Phi}_{\tau}$. Hence, we rewrite the objective function in \eqref{opt_1} as
\begin{equation} \label{AO1}
\mathcal{E}_x(\mathbf{\Phi}_{\tau}) =\mathcal{E}_x(\mathbf{\Lambda},\mathbf{\Phi}_{\tau}),
\end{equation}
as if $\mathbf{\Lambda}$ and $\mathbf{\Phi}_{\tau}$ were independent variables.
Under this hypothesis, \eqref{opt_1} can be solved by following an AO approach. The AO is an iterative algorithm whose  key advantage is that it simplifies the optimization process by breaking it into smaller subproblems, which are easier to solve. The AO approach is especially helpful when the original problem involves complicated interactions or dependencies among the variables.
In particular, we propose a two-step \emph{iterative} algorithm, where $\mathcal{E}_x$ is alternatively optimized with respect to $\mathbf{\Lambda}$ and $\mathbf{\Phi}_{\tau}$. Being $\mathbf{\Lambda}^{(k)}$ and $\mathbf{\Phi}^{(k)}_{\tau}$ the values found at iteration $k$, at iteration $k+1$ we proceed as follows: 
\begin{enumerate}
\item Having fixed the value of  $\mathbf{\Phi}_{\tau}=\mathbf{\Phi}_{\tau}^{(k)}$, we minimize
  \eqref{AO1} by computing  $\mathbf{\Lambda}^{(k+1)}$ as  $\mathbf{\Lambda}(\mathbf{\Phi}_{\tau})$ in \eqref{Lamda_Phi}. This optimization is unconstrained and  is a straightforward application of MSE minimization;  
\item Fixing $\mathbf{\Lambda}=\mathbf{\Lambda}^{(k+1)}$, the MSE takes the expression given in
\begin{align}\label{MSE_k_lambda}
&\mathcal{E}_{x}\left(\mathbf{\Lambda},\mathbf{\Phi}_{\tau}\right)=\text{tr}\left\{\mathbf{\Lambda}\mathbf{R}_y^{\mathrm{tr}}{\mathbf{\Lambda}}^{\mathrm{H}}-2\Re\left[\mathbf{\Lambda}\mathbf{\Phi}_{N\tau}\mathbf{R}_x\right]+\mathbf{R}_{x}\right\}\nonumber\\
&=\text{tr}\left\{\mathbf{\Lambda}\left(\mathbf{\Phi}_{N{\tau}}\mathbf{R}_{x}\mathbf{\Phi}_{N{\tau}}^{\mathrm{H}}+\frac{\sigma^{2}_{e}}{\rho^{\mathrm{tr}}}\mathbf{R}_{w}^{\mathrm{tr}}+\frac{\sigma^2}{\rho^{\mathrm{tr}}}\mathbf{I}_{N\tau}\right){\mathbf{\Lambda}}^{\mathrm{H}}\right.\nonumber\\
&\left.-\mathbf{\Lambda}\mathbf{\Phi}_{N\tau}\mathbf{R}_x-\mathbf{R}_x\mathbf{\Phi}_{N\tau}^{\mathrm{H}}\mathbf{\Lambda}^{\mathrm{H}}+\mathbf{R}_{x}\right\}
\end{align}
and $\mathcal{E}_x(\mathbf{\Lambda},\mathbf{\Phi}_{\tau})$ is now a \emph{convex} function of $\mathbf{\Phi}_{\tau}$. The RIS phase-shift matrix  is computed as  the solution of the minimization
\begin{equation}
\begin{aligned} \label{opt_2}
&\mathbf{\Phi}^{(k+1)}_{\tau} = \arg \min_{\mathbf{\Phi}_{\tau}\in\mathcal{F}} \mathcal{E}_x\left(\mathbf{\Lambda}^{(k+1)},\mathbf{\Phi}_{\tau}\right).
\end{aligned}
\end{equation}
\end{enumerate}
Regarding the convergence of the above algorithm we can observe that, since in both steps we minimize the MSE, at each iteration the MSE either decreases or reaches a point where it remains unchanged. Given that the MSE is a positive value, the procedure will ultimately converge to a local optimum. In the remainder of the paper, we will mention the iterations of the AO algorithm as  \emph{outer loop} iterations.

\begin{algorithm}[t]
\caption{The PG inner loop iteration}
\hspace*{0.02in} {\bf Input:}
Desirable accuracy $\varepsilon$, step size $\alpha$ and $\mathbf{\Phi}_{\tau}^{(k)}$\\
\hspace*{0.02in} {\bf Output:}
$\mathbf{\Psi}^{\star}$
\begin{algorithmic}[1]
\State {\bf Init}. initialize $s = 0$, $\mathbf{\Psi}^{(s)} =\mathbf{\Phi}^{(k)}_{\tau}$ and $\Delta_{\Psi} = 1$.
\State{\bf while} $\Delta_{\Psi}>\varepsilon$
\State$\ \ $  $\mathbf{G}^{(s)}\leftarrow \nabla_{\boldsymbol{\Psi}}\mathcal{E}_{x}\left(\mathbf{\Lambda},\mathbf{\Psi}^{(s)}\right)$ via \eqref{Gradient_2};
\State$\ \ $ $\Delta_{\Psi}\leftarrow\parallel\!\mathbf{G}^{(s)}\!\parallel$;
\State$\ \ $ $\mathbf{D}^{(s)}$ $\leftarrow$ $\left(\nabla^{2}_{\mathbf{\Psi}}\mathcal{E}_x\left(\mathbf{\Lambda},\mathbf{\Psi}^{(s)}\right)\odot \mathbf{I}_{\tau M}\right)^{-1}$ via \eqref{Hessian_2};
\State$\ \ $ $\left[\mathbf{\Psi}^{(s+1)}\right]_{i,m} \leftarrow \left[\mathbf{\Psi}^{(s)}\right]_{i,m}
-\frac{\alpha\left[\mathbf{G}^{(s)}\right]_{(m-1)\tau+i}}{\left[\mathbf{D}^{(s)} \right]_{(m-1)\tau+i,(m-1)\tau+i}}$;
\State$\ \ $ $s = s +1$; 
\State{\bf end while}
\State $\mathbf{\Psi}^{\star}\leftarrow {\boldsymbol{\Psi}}^{(s)}$;
\State \textbf{end procedure}
\end{algorithmic}\label{alg2}
\end{algorithm}

\subsubsection{Solving \eqref{opt_2} via Projected Gradient Method}
Although the objective function in  \eqref{opt_2} is convex with respect to $\mathbf{\Phi}_{\tau}$, the optimization problem is still non-convex due to the presence of the unitary modulus constraint. Given the simplicity of projecting any solution onto the feasible set $\mathcal{F}$, this type of constraint leads to the use of the \emph{projected gradient} (PG) method. The PG is  an iterative  algorithm, which, although sub-optimal, is effective and shows fast convergence towards a local optimum. The idea is to employ  gradient descent, iterated until convergence, to solve the unconstrained problem and then project the solution on the feasible set $\mathcal{F}$. In practice, we need to introduce an extra auxiliary loop variable  $\mathbf{\Psi}^{(s)}$ to describe the intermediate RIS coefficients during gradient descent. 
The application of the PG method requires to define the gradient and Hessian of the complex matrix \eqref{MSE_k_lambda}, which are given in Appendix B.

After that, the PG algorithm is initialized by setting $\mathbf{\Psi}^{(0)}={\mathbf{\Phi}}^{(k)}_{\tau}$, the last solution of the AO algorithm, then  the two steps of the PG method are:
\begin{enumerate}
\item Compute the unconstrained RIS coefficient matrix  by solving the unconstrained problem via gradient descent.  
Employing the gradient \eqref{Gradient_2} in Appendix B, $\mathbf{\Psi}^{(s)}$ is updated as
\begin{equation} \label{NewtDesc}
\!\!\left[\mathbf{\Psi}^{(s+1)}\right]_{i,m}\!\!\!\!=\left[\mathbf{\Psi}^{(s)}\right]_{i,m}\!\!\!\!-\alpha\left[\mathbf{D}^{(s)} \nabla_{\mathbf{\Psi}}\mathcal{E}_x\left(\mathbf{\Lambda},\mathbf{\Psi}^{(s)}\right)\right]_{z},
\end{equation}
where $\mathbf{D}^{(s)}\in\mathbb{C}^{\tau M\times \tau M}$ is an Hermitian positive definite matrix and the value of $\mathbf{\Lambda}$ is the outcome of the previous outer loop AO iteration. We denote with $\mathbf{\Psi}^{\star}$ the matrix obtained at  convergence.  
\item Project  $\mathbf{\Psi}^{\star}$ onto $\mathcal{F}$,  by normalizing the amplitude of each  entry to unity, i.e.,%
\begin{equation} \label{Phi_opt}
\mathbf{\Phi}_{\tau}^{(k+1)}=e^{j\angle{\mathbf{\Psi}^{\star}}}.
\end{equation}
\end{enumerate}

\begin{algorithm}[t]
\caption{Optimization of $\boldsymbol{\phi}$ during the data phase}
\hspace*{0.02in} {\bf Input:}
Desirable accuracy $\xi$, and the initial phase-shifts $\boldsymbol{\phi}^{(0)}$\\
\hspace*{0.02in} {\bf Output:}
Optimized RIS phase-shifts $\boldsymbol{\phi}_{\text{opt}}$
\begin{algorithmic}[1]
\State {\bf Init}. initialize $k = 0$ and $\Delta_{\mathcal{E}} = 1$.
\State $\mathbf{v}^{(k+1)} \leftarrow$ substitute $\boldsymbol{\phi}^{(k)}$ into \eqref{opt_2.120};
\State $\mathcal{E}_{s}({\mathbf{v}}^{(k+1)}, \boldsymbol{\phi}; \hat{\mathbf{x}}) \leftarrow$ substitute ${\mathbf{v}}^{(k+1)}$ into \eqref{MSE_s11};
\State {\bf while} $\frac{\Delta_{\mathcal{E}}}{\mathcal{E}_{s}({\mathbf{v}}^{(k+1)}, \boldsymbol{\phi}; \hat{\mathbf{x}})}>\xi$ {\bf do}
\State$\ \ $$\Delta_{\mathcal{E}}\leftarrow \mathcal{E}_{s}({\mathbf{v}}^{(k+1)}, \boldsymbol{\phi}; \hat{\mathbf{x}})$;
\State$\ \ $$\boldsymbol{\theta}^{\star}\leftarrow$ Apply the PG method for \eqref{opt_2.3};
\State$\ \ $$\boldsymbol{\phi}^{(k+1)}\leftarrow e^{j\angle{\boldsymbol{\theta}^{\star}}}$;
\State$\ \ $$k = k +1$;
\State$\ \ $${\mathbf{v}}^{(k+1)} \leftarrow$ substitute $\boldsymbol{\phi}^{(k)}$ into \eqref{opt_2.120};
\State$\ \ $$\mathcal{E}_{s}({\mathbf{v}}^{(k+1)}, \boldsymbol{\phi}; \hat{\mathbf{x}}) \leftarrow$ substitute ${\mathbf{v}}^{(k+1)}$ into \eqref{MSE_s11};
\State$\ \ $$\Delta_{\mathcal{E}}= \mathcal{E}_{s}({\mathbf{v}}^{(k+1)}, \boldsymbol{\phi}; \hat{\mathbf{x}})-\Delta_{\mathcal{E}}$;
\State {\bf end while}
\State $\boldsymbol{\phi}_{\text{opt}} \leftarrow \boldsymbol{\phi}^{(k)}$.
\State \textbf{end procedure}
\end{algorithmic}\label{alg3}
\end{algorithm}

\subsubsection{Solving \eqref{NewtDesc} via {Newton}'s Method}
Gradient descent is iterative by nature and  \eqref{NewtDesc} may need to be iterated several times before achieving convergence, so that the speed of convergence and the choice of the stepsize $\alpha$ are important issues for this type of iterative methods. In particular, when the  Hessian of the objective  function is known, we can choose
\begin{equation}
    \mathbf{D}^{(s)}=\left(\nabla^{2}_{\mathbf{\Psi}}\mathcal{E}_x\left(\mathbf{\Lambda},\mathbf{\Psi}^{(s)}\right)\right)^{-1}.
\end{equation}
In this case, the iterative algorithm is indicated as the \emph{Newton}'s method and has the great advantage of being able to find the minimum of a quadratic function as  \eqref{MSE_k_lambda} with very few iterations \cite{bertsekas2016nonlinear}. From Appendix B, for the single-input and single-output (SISO) case the Hessian is
\begin{align}\label{Hess_siso}  
\nabla^{2}_{\boldsymbol{\Psi}}\mathcal{E}_{x}\left(\mathbf{\Lambda},\mathbf{\Psi}\right) = 2\boldsymbol{\Lambda}^{\mathrm{H}}\boldsymbol{\Lambda} \otimes \mathbf{R}_c + \frac{2\sigma_{e}^2}{\rho^{\mathrm{tr}}}\left(\!\left(\boldsymbol{\Lambda}^{\mathrm{H}}\boldsymbol{\Lambda}\right)\odot\mathbf{I}_{\tau}\right) \otimes \mathbf{R}_q,
\end{align}
 and we can conclude that    the elements in the diagonal of $\nabla^{2}_{\boldsymbol{\Psi}}\mathcal{E}_{x}$ are all  positive, being obtained as products of positive factors. It can be shown that the same property applies also to the multiple-input single-output (MISO) case.  Accordingly, considered the potentially high computational complexity of inverting the $\tau M \times  \tau M$ Hessian matrix, a simplified version of the Newton's method, which is valid when the elements on the diagonal of the Hessian are all strictly positive,  is obtained by approximating the Hessian by the elements of its main diagonal, so that it is    
\begin{equation}
    \mathbf{D}^{(s)}=\left(\nabla^{2}_{\mathbf{\Psi}}\mathcal{E}_x\left(\mathbf{\Lambda},\mathbf{\Psi}^{(s)}\right)\odot \mathbf{I}_{\tau M}\right)^{-1}.
\end{equation}
In this specific case, the update rule for the \emph{diagonally scaled steepest descent} method takes the form
\begin{equation} \label{Phi_bound}
\!\left[\mathbf{\Psi}^{(s+1)}\right]_{i,m}\!\!\!=\!\left[\mathbf{\Psi}^{(s)}\right]_{i,m}
\!\!-\frac{\alpha\left[\nabla_{\mathbf{\Psi}}\mathcal{E}_{x}\right]_{(m-1)\tau+i}}{\left[\nabla^{2}_{\mathbf{\Psi}}\mathcal{E}_{x} \right]_{(m-1)\tau+i,(m-1)\tau+i}}.
\end{equation}
The AO outer and PG inner loop iterations for obtaining $\mathbf{\Phi}^{\text{opt}}_{\tau}$ are summarized in Algorithm \ref{alg1} and Algorithm \ref{alg2}, respectively. 

\subsection{Optimizing RIS for Data Transmission}\label{OPT_COMM}


The optimization of the RIS during the data transmission phase follows the same steps. To proceed further, we rewrite \eqref{MSE_MMSE_s1_1} as a function of $\mathbf{v}$ and $\boldsymbol{\phi}$ and formulate the optimization problem as follows:
\begin{equation}
\begin{aligned} \label{opt_2.12}
&\min_{\boldsymbol{\phi}}\ \mathcal{E}_{s}(\mathbf{v},\boldsymbol{\phi}; \hat{\mathbf{x}}), \ \ \! \textrm{s.t.}\ \ |[\boldsymbol{\phi}]_{m}|=1.
\end{aligned}
\end{equation}
Considered that also $\mathbf{R}_y$ depends on $\boldsymbol{\phi}$,  $\mathcal{E}_{s}(\mathbf{v},\boldsymbol{\phi}; \hat{\mathbf{x}})$ is a non-convex function of $\boldsymbol{\phi}$ and to solve \eqref{opt_2.12} we can follow once again the iterative  \emph{AO} approach as illustrated in Algorithm \ref{alg3}. 

Let $\boldsymbol{\phi}^{(k)}$ be the vector of RIS  coefficients at iteration $k$, then  $\mathbf{v}^{(k+1)}$ is computed by replacing $\boldsymbol{\Phi}_N$ with $\mathbf{I}_N\otimes{\boldsymbol{\phi}^{(k)}}^{\mathrm{T}}$ in \eqref{opt_2.120}. The new vector of RIS coefficients $\boldsymbol{\phi}^{(k+1)}$ can be found by solving 
\begin{align} \label{opt_2.3}
&\boldsymbol{\phi}^{(k+1)} = \arg \min_{\boldsymbol{\phi}}\ \mathcal{E}_{s}(\mathbf{v}^{(k+1)},\boldsymbol{\phi}; \hat{\mathbf{x}}),\nonumber\\
&\qquad\qquad \textrm{s.t.}\ \ |[\boldsymbol{\phi}]_{m}|=1,
\end{align}
and it can be solved by applying the PG method, with the required gradient and Hessian provided in \eqref{G_MSE_2_com} and \eqref{Hessian_s_2} of Appendix B.  
Since the MSE decreases at each iteration until it reaches a point where it remains unchanged, the procedure necessarily converges to a local optimum. 
After that, by substituting $\mathbf{v}(\boldsymbol{\phi}_{\text{opt}})$ into \eqref{SE_2}, the achievable SE of the system with optimal MMSE design can be obtained.

\begin{figure}[t]
\setlength{\abovecaptionskip}{-0.08cm}
\setlength{\belowcaptionskip}{-0.1cm}   
\begin{center}
\includegraphics[scale=0.540]{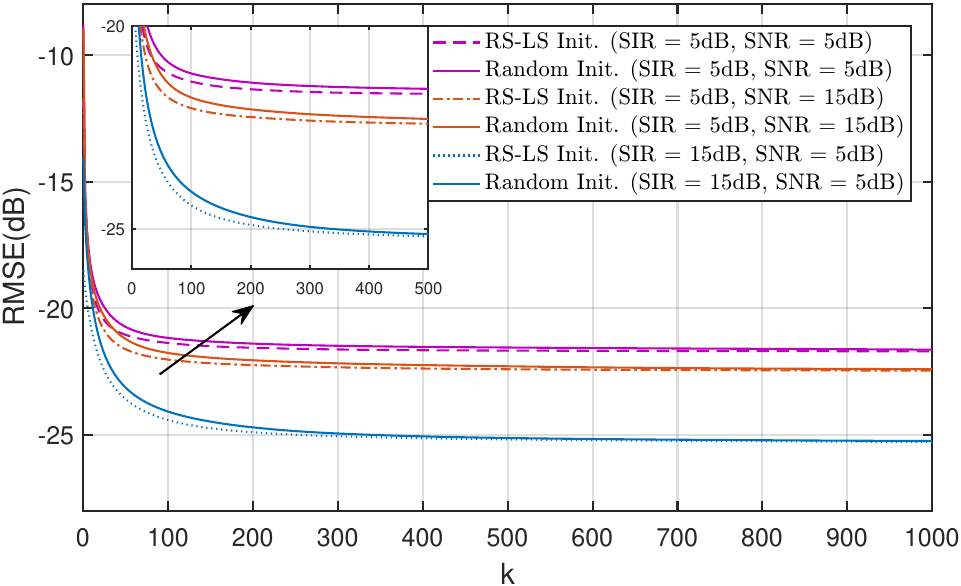}
\end{center}
\caption{The RMSEs vs. $k$ for the LMMSE estimator with AO.}
\label{FigP8}
\end{figure}

\begin{table}[b]
\caption{Simulation parameters.}
\begin{center}
\begin{tabular}{|cccccc|}
\hline
\multicolumn{4}{|c|}{Cascaded Channel}                                                                                                            & \multicolumn{2}{c|}{EMI}                          \\ \hline
\multicolumn{1}{|c|}{$\varphi_h$}  & \multicolumn{1}{c|}{$\vartheta_h$}    & \multicolumn{1}{c|}{$\varphi_{g_n}$}   & \multicolumn{1}{c|}{$\vartheta_{g_n}$}    & \multicolumn{1}{c|}{$\varphi_e$}   & $\vartheta_e$   \\ \hline
\multicolumn{1}{|c|}{$70^{\circ}$} & \multicolumn{1}{c|}{$-20^{\circ}$} & \multicolumn{1}{c|}{$-60^{\circ}$} & \multicolumn{1}{c|}{$-30^{\circ}$} & \multicolumn{1}{c|}{$-10^{\circ}$} & $20^{\circ}$ \\ \hline
\multicolumn{2}{|c|}{$\triangle_h = 10^{\circ}$}                                                 & \multicolumn{2}{c|}{$\triangle_{g_n} = 5^{\circ}$}                                                  & \multicolumn{2}{c|}{$\triangle_e = 20^{\circ}$}                            \\ \hline
\multicolumn{6}{|c|}{$\{f_i(\varphi,\vartheta)|i=g_n,h,e\}$: Gaussian Model}                                                                                                                           \\ \hline
\end{tabular}
\end{center}
\label{table1}
\end{table}

\section{Numerical Results}

\begin{figure}[t]  
\centering
\setlength{\abovecaptionskip}{-0.1cm}
\setlength{\belowcaptionskip}{-0.3cm} 
\subfigure[$\!\!\!\!\!\!\!\!\!\!\!\!\!\!\!\!\!\!$]{
\includegraphics[scale=0.528]{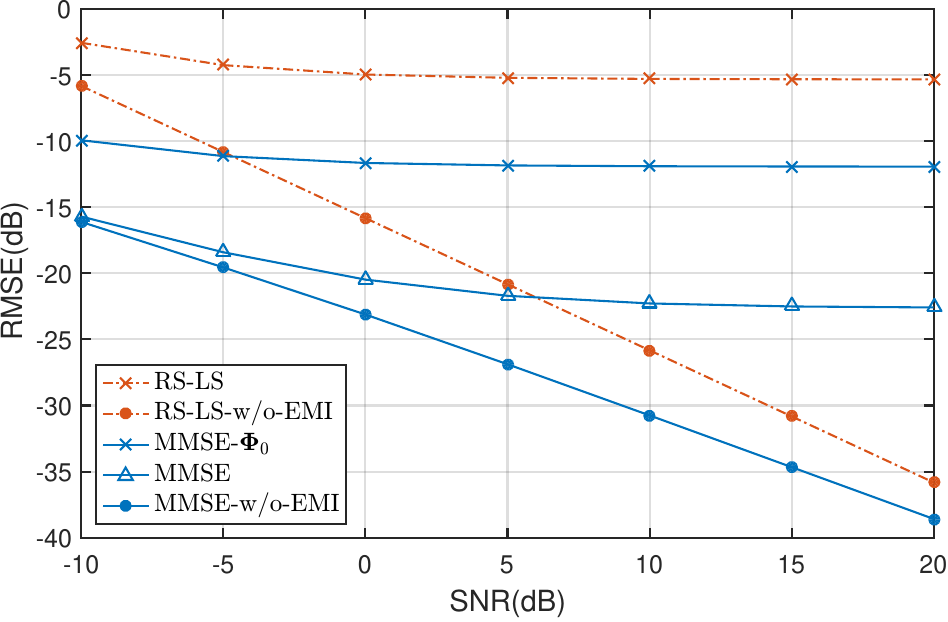}}\\
\subfigure[$\!\!\!\!\!\!\!\!\!\!\!\!\!\!\!\!\!\!$]{
\includegraphics[scale=0.530]{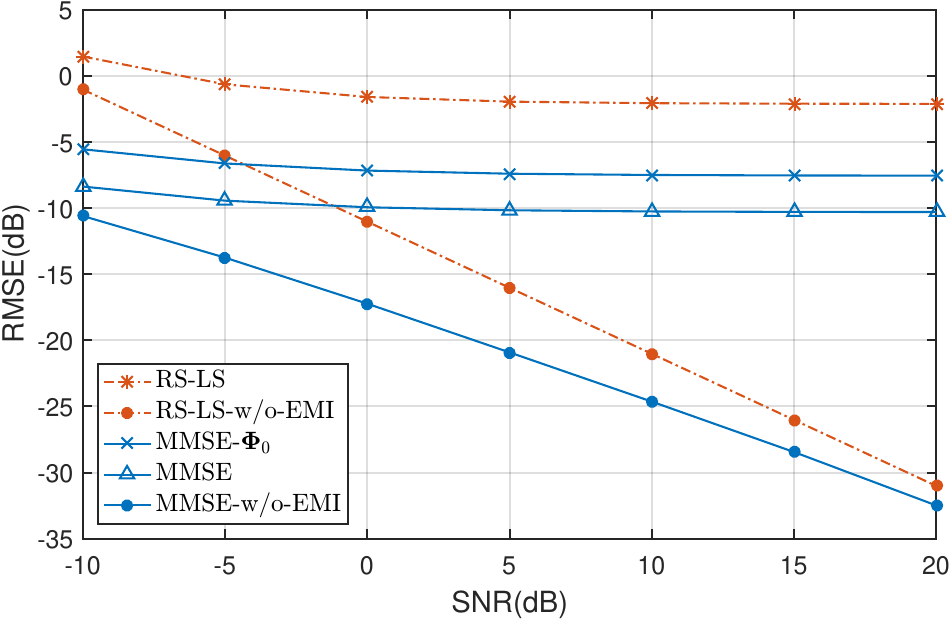}}
\caption{The RMSEs vs. SNR for different estimators and RIS phase-shift configurations. (a) SISO case. (b) MISO case.} 
\label{FigP1}
\end{figure}

Numerical results are now given to verify the performance of the proposed channel estimation and transmission schemes. Taking simulation speed into consideration, we primarily test the proposed solution in the SISO scenario, with the exception of Fig. \ref{FigP1} (b). The required gradient and Hessian matrices in the SISO case are provided in Appendix B. Unless otherwise stated, the RIS is equipped with $M = 36$ elements with $M_H = M_V = 6$, and the vertical and horizontal inter-element distances are set to $\lambda/2$, where $\lambda$ is the wavelength. The scattering function $f_h(\varphi,\vartheta)$ in the spatial correlation matrix $\mathbf{R}_h$ for the UE-RIS channel follows the Gaussian distribution within the $\triangle_h = 10^\circ$ neighborhood of $(\varphi_h,\vartheta_h) = (70^\circ,-20^\circ)$, and the scattering function $f_{g_n}(\varphi,\vartheta)$ in $\mathbf{R}_{g_n}$ also follows the Gaussian distribution within the $\triangle_{g_n} = 5^\circ$ neighborhood of $(\varphi_{g_n},\vartheta_{g_n}) = (-60^\circ,-30^\circ)$. The scattering function $f_e(\varphi,\vartheta)$ in $\mathbf{R}_e$ for the EMI is selected the Gaussian distribution within the $\triangle_e = 20^\circ$ neighborhood of $(\varphi_e,\vartheta_e) = (-10^\circ,20^\circ)$, as illustrated in Table \ref{table1}. The pilot length is set to $\tau = M = 36$ and the coherence interval is $\tau_c = 10\tau = 360$ symbols. 

\subsection{Channel Estimation Analysis}

\begin{figure}[t]
\setlength{\abovecaptionskip}{-0.2cm}
\setlength{\belowcaptionskip}{-0.4cm}   
\begin{center}
\includegraphics[scale=0.55]{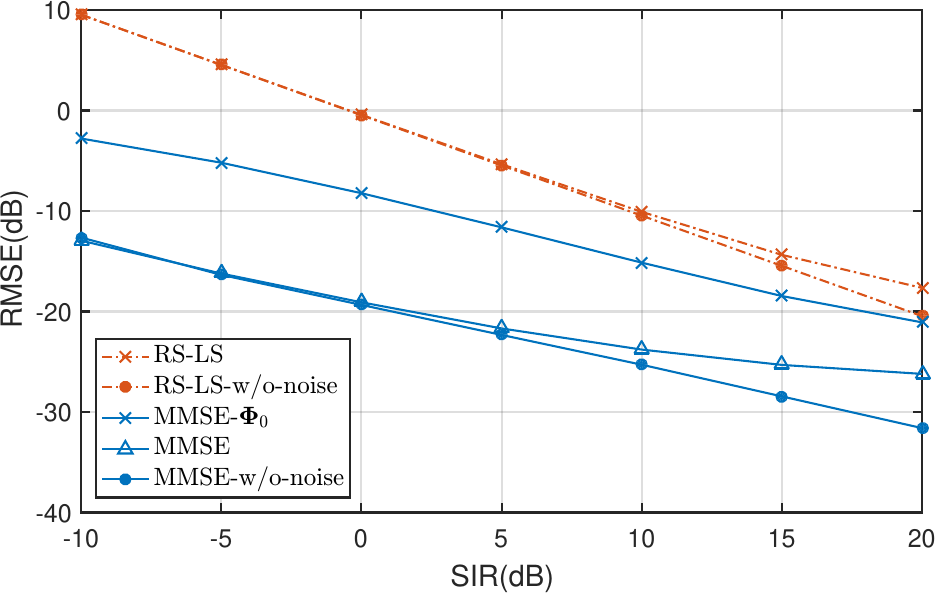}
\end{center}
\caption{The RMSEs vs. SIR for different estimators and RIS phase-shift configurations.}
\label{FigP2}
\end{figure}

\begin{figure}[t]
\setlength{\abovecaptionskip}{-0.2cm}
\setlength{\belowcaptionskip}{-0.15cm}   
\begin{center}
\includegraphics[scale=0.54]{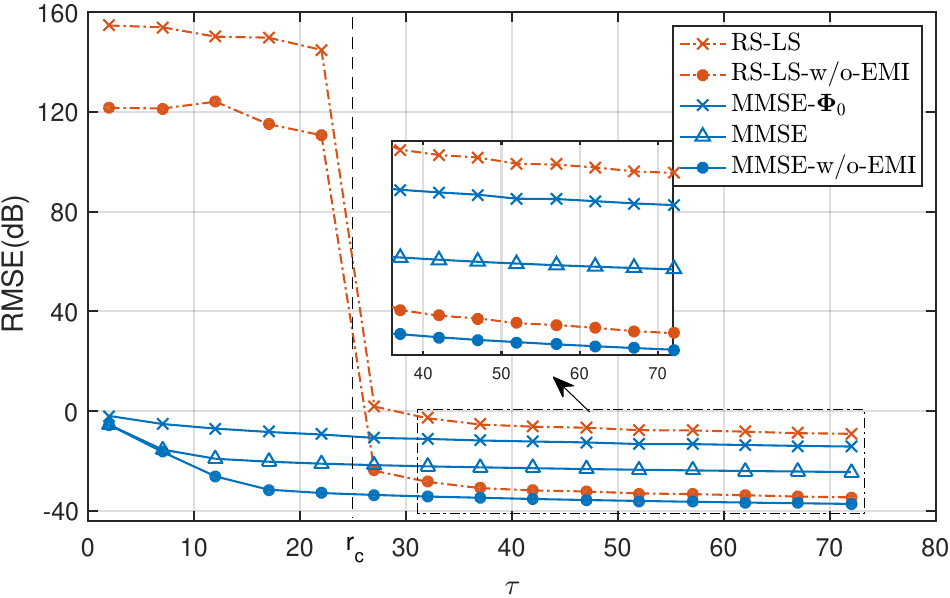}
\end{center}
\caption{The RMSEs vs. $\tau$ for different estimators and RIS phase-shift configurations.}
\label{FigP4}
\end{figure}

\begin{figure}[t]
\setlength{\abovecaptionskip}{-0.15cm}
\setlength{\belowcaptionskip}{-0.455cm}   
\begin{center}
\includegraphics[scale=0.53]{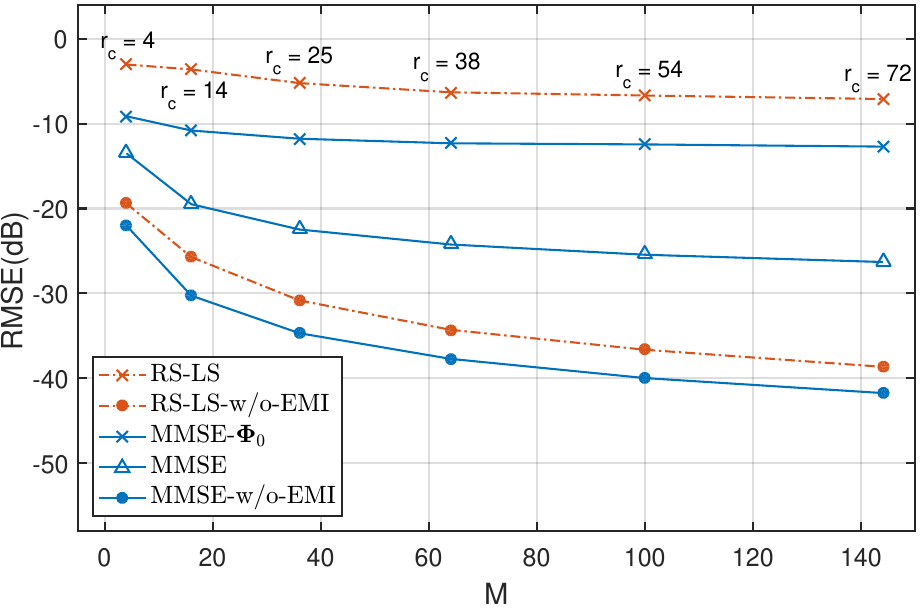}
\end{center}
\caption{The RMSEs vs. $M$ for different estimators and RIS phase-shift configurations.}
\label{FigP5}
\end{figure}

\begin{figure}[t]
\setlength{\abovecaptionskip}{-0.15cm}
\setlength{\belowcaptionskip}{-0.6cm}   
\begin{center}
\includegraphics[scale=0.54]{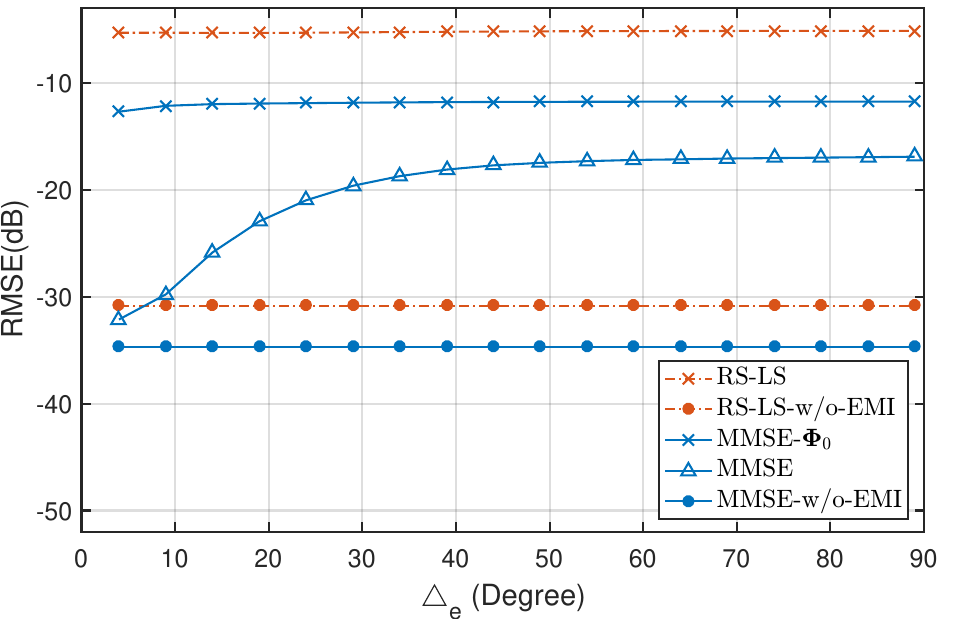}
\end{center}
\caption{The RMSEs vs. $\triangle_e$ for different estimators and RIS phase-shift configurations.}
\label{FigP3}
\end{figure}

In Fig. \ref{FigP8} to Fig. \ref{FigP3}, the performance in terms of the relative MSE (RMSE) of the channel estimation, defined as $\mathbb{E}\{{\|\mathbf{x}-\hat{\mathbf{x}}\|^2}/{M}\}$, is shown for three different approaches. The first is the proposed MMSE optimization approach with AO, defined in Algorithms \ref{alg1} and \ref{alg2} and simply referred to as `MMSE' in the figures. The second is the RS-LS estimator defined in \eqref{RSLS_estimator}, where the RIS is optimized based on \cite{Demir2022Exploiting}. The RIS configuration obtained by RS-LS is denoted as $\mathbf{\Phi}_0$ and used as the initial phase of the iterative MMSE scheme in Algorithm \ref{alg1}. The third approach is referred to as `MMSE-$\mathbf{\Phi}_0$', where the RIS shifts are $\mathbf{\Phi}_0$, but unlike RS-LS, the linear filter is calculated as in \eqref{Lamda_Phi}, i.e., based on the LMMSE criterion. For the MMSE approach, we assume $\alpha = 0.5$ and $\epsilon = 10^{-5}$, and to improve simulation speed, we apply the diagonally scaled steep descent method only once within the inner loop iteration defined in Algorithm \ref{alg2}.

%

Fig. \ref{FigP8} depicts the RMSE of the LMMSE estimator with the AO algorithm as a function of the outer iteration count $k$ under varying SNRs and SIRs. The abbreviation `RS-LS Init.' signifies that the AO algorithm is initiated with $\mathbf{\Phi}_0$, the optimal phase-shifts for the RS-LS estimator. In contrast, `Random Init.' denotes the initialization of the AO algorithm with random unit-modulus phase-shifts. As detailed in Section IV, the AO algorithm converges progressively as $k$ increases. Notably, the AO algorithm exhibits slightly faster convergence when initialized with $\mathbf{\Phi}_0$ compared to when initialized with random unit-modulus phase-shifts. This observation motivates our choice to employ the $\mathbf{\Phi}_0$ initialization in the forthcoming simulations.

\begin{figure}[t]
\setlength{\abovecaptionskip}{-0.0cm}
\setlength{\belowcaptionskip}{0.0cm}   
\centering
\includegraphics[scale=0.525]{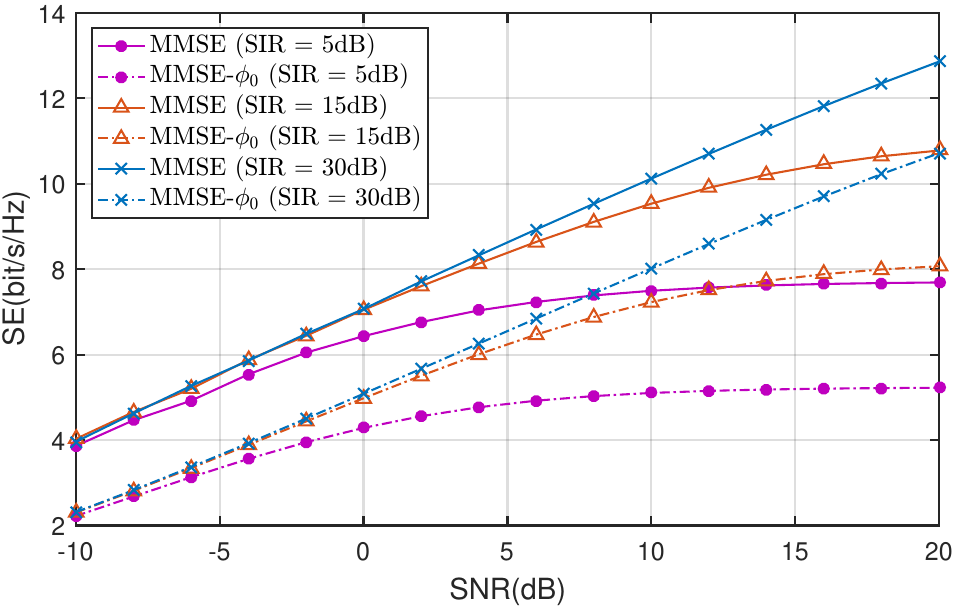}
\caption{The SE vs. SNR for the MMSE receiver.}
\label{FigP7}
\end{figure}
\begin{figure}[t]
\setlength{\abovecaptionskip}{-0.1cm}
\setlength{\belowcaptionskip}{-0.2cm}   
\centering
\includegraphics[scale=0.51]{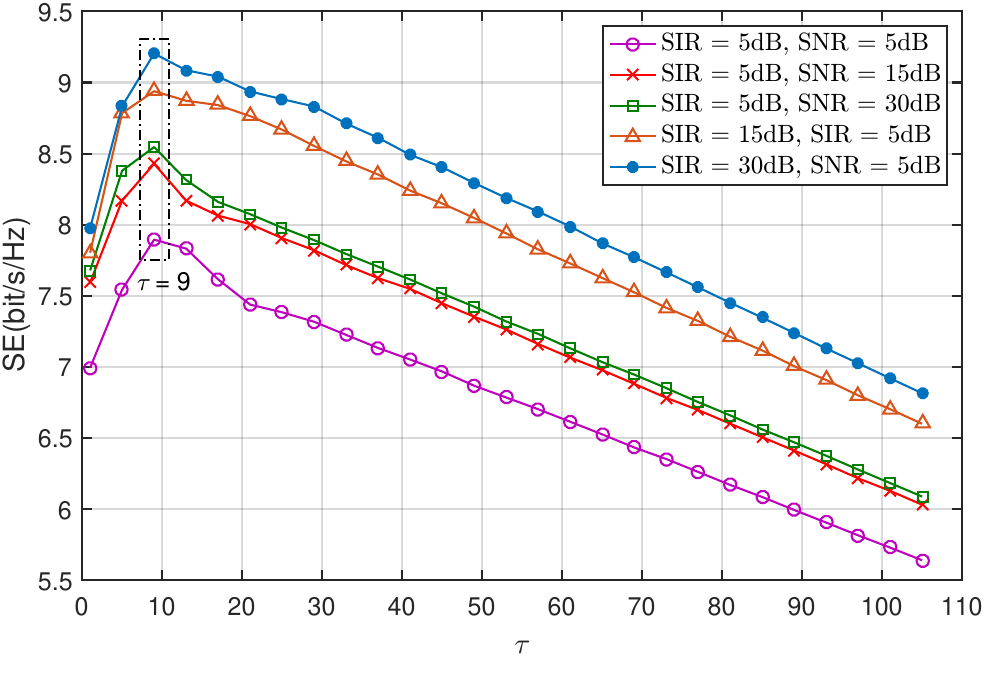}
\caption{The SE vs. $\tau$ for the MMSE receiver.}
\label{FigP6}
\end{figure}

Fig. \ref{FigP1} shows the RMSE as a function of SNR when SIR = 5dB both in the SISO and MISO scenarios. In the MISO case, the UPA at the BS consists of $2\times 2$ antennas. For comparison purposes, we also give the results of MMSE and RS-LS schemes when SIR $= \infty$, i.e. EMI is not present, which are denoted `w/o EMI'.
Let us first consider the case where the RIS is not affected by interference. In this case, both the LMMSE and RS-LS estimators show linearly decreasing MSE with SNR. Also, LMMSE outperforms RS-LS especially at low SNRs, which is consistent with expectations for MMSE estimation compared to LS. In the presence of EMI, a threshold approximately corresponding to SNR = SIR is observed in all cases, i.e., when SIR begins to be the predominant effect. Moreover, `MMSE-$\mathbf{\Phi}_0$' outperforms the RS-LS method by almost $5$ dB, even though both methods use the same RIS configuration $\mathbf{\Phi}_0$. This is because, in the MMSE case, the LMMSE criterion takes into account both noise and EMI statistics. More importantly, the proposed MMSE scheme provides consistent gain in terms of `MMSE-$\mathbf{\Phi}_0$', which proves the effectiveness of the proposed iterative RIS optimization approach in Algorithms \ref{alg1} and \ref{alg2}. Finally, comparing Fig. \ref{FigP1} (a) and Fig. \ref{FigP1} (b), in the MISO scenario, all estimators exhibit consistency in performance with those in the SISO scenario. However, due to the increased channel dimension that needs to be estimated, their overall performance is slightly inferior to the estimators in the SISO scenario.

Fig. \ref{FigP2} shows the RMSE as a function of SIR at an SNR of 5 dB. The performance of all algorithms shows similar behavior as in the previous case, with the difference that the threshold appears for SIR $\gg$ SNR, indicating that EMI is the most detrimental effect for channel estimation. As before, the best results are obtained with the proposed LMMSE estimator, and this behavior is more evident for low SIR, i.e., when the impact of EMI is higher. Specifically, the performance gap between RS-LS and MMSE at SIR $= 20$ dB is of $8.55$ dB, whereas the gap increases to $18.65$ dB at SIR $= 0$ dB. 

Fig. \ref{FigP4} shows the RMSE as a function of the pilot length $\tau$ when SNR = 15dB and SIR = 5dB. We show in the figure $r_c = \text{rank}\{\mathbf{R}_{c}\}$, which is the minimum number of pilots that can be used by the RS-LS scheme according to \cite{Demir2022Exploiting}. It can be seen from the figure that the LMMSE estimator can be applied for any $\tau \geq 1$, but naturally improves its performance as $\tau$ increases. The largest performance improvement is observed for $\tau < r_c$, while the performance tends to reach a threshold as $\tau$ approaches the number of RIS elements $M$. Conversely, as expected, the performance of the RS-LS estimator for $\tau < r_c$ is very poor due to the insufficient number of pilots. Also in this case, the performance tends to stabilize as $\tau$ approaches $M$.

Fig. \ref{FigP5} shows the RMSE as a function of $M$ at SNR = 15dB and SIR = 5dB. Note that increasing $M$ increases the number of pilots to be estimated. However, this has only a limited effect on the channel rank, i.e., on the effective channel dimension. Thus, the main effect is to increase the RIS gain, which is beneficial for channel estimation. Accordingly, the RMSE value of both estimators decreases as $M$ increases. It is also worth noting that the difference in performance between MMSE and RS-LS increases with $M$.

So far, we have assumed that the  interference and the pilot signals come from different directions and do not overlap. In Fig. \ref{FigP3}, we evaluate the RMSE as a function of angular spread, considering $\triangle_e$ from $4^\circ$ to $89^\circ$. The SNR and SIR are both set to 15 dB. As expected, the gap between `MMSE' and `MMSE-w/o-EMI' increases as $\triangle_e$ increases. When the interference approaches isotropic scattering, i.e., $\triangle_e \geq 50^\circ$, the gap between the two curves reaches $17.34$ dB. This is due to the physical overlap between the UE-RIS channel $\mathbf{h}$ and the interference $\mathbf{e}(i)$. Since only the signal subspace under the assumption of isotropic conditions is used to design $\mathbf{\Phi}_0$ \cite{Demir2022Exploiting}, the estimators `MMSE-$\mathbf{\Phi}_0$' and `RS-LS' are less sensitive to changes in $\triangle_e$ compared to the `MMSE' estimator. However, even in the isotropic case, the MMSE estimator still retains a performance advantage of about $11.78$ dB over the RS-LS estimator.

\subsection{Spectral Efficiency Analysis}

Fig. \ref{FigP7} presents the SE as a function of SNR at various SIRs. We consider two approaches: the first one is the MMSE combiner with AO defined in Algorithm \ref{alg3}, labeled as `MMSE'. The initial $\boldsymbol\phi_0$ for the AO algorithm is set to a row of $\boldsymbol{\Phi}^{\text{opt}}_{\tau}$, the optimal phase-shifts for the LMMSE estimator. Additionally, the MMSE combiner with $\boldsymbol\phi_0$ is provided in Fig. \ref{FigP7} as a benchmark. When the RIS is almost unaffected by white interference, i.e., SIR $= 30$dB, the SE of all combiners exhibits nearly linear growth with increasing SNR. In the presence of EMI, a threshold slightly above SNR $=$ SIR can be observed in all combiners. This indicates that the EMI is the most adverse factor inhibiting SE in the data transmission phase. Furthermore, after employing the AO algorithm, the performance of the MMSE combiner is improved by over $2$ bits/s/Hz compared to `MMSE-$\boldsymbol\phi_0$'. This underlines the effectiveness of the AO algorithm in the transmission part.

Fig. \ref{FigP6} shows the SE as a function of pilot length at different SNRs and SIRs, where the pilot length is set in the range 1 to $\tau_{\text{max}} = 3M =108$. We see that the SE does not continuously increase with the increase of the pilot length. When the allocated pilot length is sufficient to accurately estimate $\mathbf{\hat{x}}$, i.e., $\tau > 9$ in the simulated system, additional pilot sequences would consume the avaiable coherence interval $\tau_c$, leading to a decrease in the SE. Therefore, a compromise design of the pilot length is crucial for achieving the optimal SE in RIS-aided communications, and this is left for our future work.

\section{Conclusions}

This paper is centered on a RIS-aided system in which a single-antenna UE exchanges information with a multi-antenna BS, and addressed the problem of channel estimation and SE optimization. Both problems were formulated aiming at the minimization of the MSE. A key aspect is the concurrent optimization of the RIS coefficients with spatially correlated channels and EMI. The ensuing optimization tasks yield two non-convex problems, both successfully addressed through the application of an iterative algorithm based on the principle of alternating optimization, with demonstrated convergence towards locally optimal solutions. Numerical results substantiate the efficacy of the proposed method, showcasing its superiority over state-of-the-art alternatives.

\section*{Appendix A}

We assume that the receiver has knowledge of the statistics of $\mathbf{x}$ and $\mathbf{w}^{\mathrm{tr}}$ so that the LMMSE estimate of $\mathbf{{x}}$ can be computed.  
The spatial correlation matrix $\mathbf{R}_{x}\in\mathbb{C}^{N M\times N M}$ of $\mathbf{{x}}$ is 
\begin{align}\label{App_Rx}
\mathbf{R}_{x}= \mathbb{E}\{\mathbf{x}\mathbf{x}^{\mathrm{H}}\}=\mathbf{R}_{g'_m}\otimes\left(\mathbf{R}_{g_n}\odot\mathbf{R}_{h}\right)= \mathbf{R}_{g'_m} \otimes\mathbf{R}_{c},
\end{align}
from the statistical independence of $\mathbf{h}$ and $\{\mathbf{g}_n\}$. 

Since the interference $\{\mathbf{e}({i});i=1,2,\ldots, \tau\}$ is spatially correlated but uncorrelated in time, we have
\begin{align}
\begin{split}
\mathbb{E}\left\{w_{n_1}(i) {w_{n_2}(j)}^{*}\right\}\! = \!
\left \{\!
\begin{array}{ll}
   \sigma_e^2 \mathbf{R}_{g'_m}({n_1},{n_2})\bm{\phi}(i)^{\mathrm{T}}\mathbf{R}_{q}\bm{\phi}(i)^*\! \!\!                 & i = j,\\
    0,  \!  \! \!                        & i\neq j,\\
\end{array}
\right.
\end{split}
\end{align}
where $\mathbf{R}_{g'_m}({n_1},{n_2})$ is the $({n_1},{n_2})$-th element of the correlation matrix $\mathbf{R}_{g'_m}$.  
Hence, the correlation matrix of $\mathbf{w}^{\mathrm{tr}}$ is given by
\begin{align}\label{App_Rw_prime}
\mathbf{R}_{w}^{\mathrm{tr}} = \frac{1}{\sigma^{2}_{e}}\mathbb{E}\{\mathbf{w}^{\mathrm{tr}}{(\mathbf{w}^{\mathrm{tr}})}^{\mathrm{H}}\}
=\mathbf{R}_{g'_m}\otimes\left(\!\left(\mathbf{\Phi}_{\tau}\mathbf{R}_{q}\mathbf{\Phi}_{\tau}^{\mathrm{H}}\right)\odot\mathbf{I}_{\tau}\right).
\end{align}
The LMMSE estimator is the linear filter designed to minimize the MSE between the channel $\mathbf{x}$ and its MMSE estimate $\hat{\mathbf{{x}}} = \frac{1}{\sqrt{\rho^{\mathrm{tr}}}}\mathbf{R}_{x}\mathbf{\Phi}_{N{\tau}}^{\mathrm{H}}\left({{\bf R}_y^{\mathrm{tr}}}\right)^{-1}\mathbf{y}^{\mathrm{tr}}$, i.e., 
\begin{align}\label{App_x_MMSE}
\mathcal{E}_x(\mathbf{\Phi}_{{\tau}}) = \mathbb{E}\left\{\left\|\mathbf{x}-\hat{\mathbf{x}}\right\|^2\right\}
=\text{tr}\left\{\mathbf{R}_{x}-\mathbf{R}_x\mathbf{Q}^{\mathrm{tr}}(\mathbf{\Phi}_{{\tau}})\mathbf{R}_x\right\}.
\end{align}

\section*{Appendix B \\ Gradient and Hessian of \eqref{MSE_k_lambda} and \eqref{MSE_s11}}\label{AppB}

Let $f:\mathbb{C}^{P\times Q}\rightarrow \mathbb{R}$ be a function that is twice differentiable, we define the complex \emph{gradient} operator as the $PQ$-dimensional vector
\begin{equation}\label{gradVect}
\nabla_{\mathbf{X}}f = \frac{\partial f}{\partial \mathbf{x}^{*}}
\end{equation} 
where $\mathbf{x} = \vvec(\mathbf{X})$. Therefore, if we let $z =(q-1)Q+p$, then it follows that $\left[\nabla_{\mathbf{X}}f \right]_{z}=\frac{\partial f}{\partial \mathbf{x}^{*}_{z}}=\frac{\partial f}{\partial \mathbf{X}^{*}_{p,q}}$.

Considering the fact that $\mathbf{R}_{x}=\mathbf{R}_{g'_m}\otimes\mathbf{R}_{c}$, $\mathbf{R}^{\mathrm{tr}}_{w}=\mathbf{R}_{g'_m}\otimes\left(\!\left(\mathbf{\Phi}_{\tau}\mathbf{R}_{q}\mathbf{\Phi}_{\tau}^{\mathrm{H}}\right)\odot\mathbf{I}_{\tau}\right)$ and $\mathbf{\Phi}_{N\tau}$ $=$ $\mathbf{I}_N\otimes\mathbf{\Phi}_{\tau}$, we extract the terms in \eqref{MSE_k_lambda} that are related to $\boldsymbol{\Phi}_{\tau}$, and represent them sequentially as follows:
\begin{align} \label{E_1}
\mathcal{E}_{1}^{(k+1)} = \text{tr}\big\{{\mathbf{\Lambda}^{(k+1)}}^{\mathrm{H}}\mathbf{\Lambda}^{(k+1)}\underbrace{\mathbf{R}_{g'_m}\otimes\mathbf{\Phi}_{\tau}\mathbf{R}_c\mathbf{\Phi}_{\tau}^{\mathrm{H}}}_{\mathbf{A}_{\mathbf{\Phi}_{\tau}}}\big\},
\end{align}
\begin{align} \label{E_2}
\mathcal{E}_{2}^{(k+1)} = \frac{\sigma_{e}^2}{{\rho^{\mathrm{tr}}}}\text{tr}\big\{{\mathbf{\Lambda}^{(k+1)}}^{\mathrm{H}}\mathbf{\Lambda}^{(k+1)}\underbrace{\mathbf{R}_{g'_m}\otimes\left[\left(\mathbf{\Phi}_{\tau}\mathbf{R}_q\mathbf{\Phi}_{\tau}^{\mathrm{H}}\right)\odot\mathbf{I}_{\tau}\right]}_{\mathbf{B}_{\mathbf{\Phi}_{\tau}}}\big\},
\end{align}
\begin{align} \label{E_3}
\mathcal{E}_{3}^{(k+1)} = \text{tr}\big\{\mathbf{\Lambda}^{(k+1)}\underbrace{\mathbf{R}_{g'_m}\otimes\mathbf{\Phi}_{\tau}\mathbf{R}_c}_{\mathbf{C}_{\mathbf{\Phi}_{\tau}}}\big\},
\end{align}
and
\begin{align} \label{E_4}
\mathcal{E}_{4}^{(k+1)} = \text{tr}\big\{{\mathbf{\Lambda}^{(k+1)}}^{\mathrm{H}}\underbrace{\mathbf{R}_{g'_m}\otimes \mathbf{R}_c\mathbf{\Phi}_{\tau}^{\mathrm{H}}}_{\mathbf{D}_{\mathbf{\Phi}_{\tau}}}\big\}.
\end{align}
\vspace{-0.0cm}
Applying the chain rule and abbreviating $\mathbf{\Lambda}^{(k+1)}$ to $\mathbf{\Lambda}$, the gradient of \eqref{E_1} to \eqref{E_4} with respect to $[\mathbf{\Phi}_{\tau}]_{i,m}$ can be expressed as
\begin{align} \label{g_E_1}
&\nabla_{[\mathbf{\Phi}_{\tau}]_{i,m}}\mathcal{E}_{1}^{(k+1)} = \text{tr}\left\{\left[\frac{\partial\mathcal{E}_{1}^{(k+1)}}{\partial\mathbf{A}_{\mathbf{\Phi}_{\tau}}}\right]^{\mathrm{H}}\frac{\partial\mathbf{A}_{\mathbf{\Phi}_{\tau}}}{\partial[\mathbf{\Phi}_{\tau}]_{i,m}}\right\} \nonumber\\
&= \text{tr}\left\{{\mathbf{\Lambda}}^{\mathrm{H}}\mathbf{\Lambda}\left[\mathbf{R}_{g'_m}\otimes\frac{\left(\partial\mathbf{\Phi}_{\tau}\right)\mathbf{R}_c\mathbf{\Phi}_{\tau}^{\mathrm{H}}+\mathbf{\Phi}_{\tau}\mathbf{R}_c\left(\partial\mathbf{\Phi}_{\tau}\right)^{\mathrm{H}}}{\partial[\mathbf{\Phi}_{\tau}]_{i,m}}\right]\right\}\nonumber\\
&=\text{tr}\left\{{\mathbf{\Lambda}}^{\mathrm{H}}\mathbf{\Lambda}\left[\mathbf{R}_{g'_m}\otimes\left(\mathbf{J}_{i,m}\mathbf{R}_{c}\mathbf{\Phi}_{\tau}^{\mathrm{H}}+\mathbf{\Phi}_{\tau}\mathbf{R}_{c}\mathbf{J}_{i,m}^{\mathrm{H}}\right)\right]\right\},
\end{align}
\begin{align} \label{g_E_2}
&\nabla_{[\mathbf{\Phi}_{\tau}]_{i,m}}\mathcal{E}_{2}^{(k+1)} = \frac{\sigma_{e}^2}{\rho^{\mathrm{tr}}}\text{tr}\left\{\left[\frac{\partial\mathcal{E}_{2}^{(k+1)}}{\partial\mathbf{B}_{\mathbf{\Phi}_{\tau}}}\right]^{\mathrm{H}}\frac{\partial\mathbf{B}_{\mathbf{\Phi}_{\tau}}}{\partial[\mathbf{\Phi}_{\tau}]_{i,m}}\right\}= \frac{\sigma_{e}^2}{\rho^{\mathrm{tr}}}\cdot\nonumber\\
&\text{tr}\left\{{\mathbf{\Lambda}}^{\mathrm{H}}\mathbf{\Lambda}\!\left[\mathbf{R}_{g'_m}\otimes\left(\frac{\left(\partial\mathbf{\Phi}_{\tau}\right)\mathbf{R}_q\mathbf{\Phi}_{\tau}^{\mathrm{H}}\!+\!\mathbf{\Phi}_{\tau}\mathbf{R}_q\left(\partial\mathbf{\Phi}_{\tau}\right)^{\mathrm{H}}}{\partial[\mathbf{\Phi}_{\tau}]_{i,m}}\!\odot\!\mathbf{I}_{\tau}\!\right)\!\right]\!\right\}\nonumber\\
&=\frac{\sigma_{e}^2}{\rho^{\mathrm{tr}}}\text{tr}\left\{{\mathbf{\Lambda}}^{\mathrm{H}}\mathbf{\Lambda}\left[\mathbf{R}_{g'_m}\!\otimes\!\left(\left(\mathbf{J}_{i,m}\mathbf{R}_{q}\mathbf{\Phi}_{\tau}^{\mathrm{H}}+\mathbf{\Phi}_{\tau}\mathbf{R}_{q}\mathbf{J}_{i,m}^{\mathrm{H}}\right)\!\odot\!\mathbf{I}_{\tau}\!\right)\!\right]\!\right\},
\end{align}
\begin{align} \label{g_E_3}
&\nabla_{[\mathbf{\Phi}_{\tau}]_{i,m}}\mathcal{E}_{3}^{(k+1)} = \text{tr}\left\{\left[\frac{\partial\mathcal{E}_{3}^{(k+1)}}{\partial\mathbf{C}_{\mathbf{\Phi}_{\tau}}}\right]^{\mathrm{H}}\frac{\partial\mathbf{C}_{\mathbf{\Phi}_{\tau}}}{\partial[\mathbf{\Phi}_{\tau}]_{i,m}}\right\}\nonumber\\
&= \text{tr}\left\{\mathbf{\Lambda}\left(\mathbf{R}_{g'_m}\otimes\frac{\left(\partial\mathbf{\Phi}_{\tau}\right)\mathbf{R}_c}{\partial[\mathbf{\Phi}_{\tau}]_{i,m}}\right)\right\}
=\text{tr}\left\{{\mathbf{\Lambda}}\left(\mathbf{R}_{g'_m}\otimes\mathbf{J}_{i,m}\mathbf{R}_{c}\right)\right\},
\end{align}
and
\begin{align} \label{g_E_4}
&\nabla_{[\mathbf{\Phi}_{\tau}]_{i,m}}\mathcal{E}_{4}^{(k+1)} = \text{tr}\left\{\left[\frac{\partial\mathcal{E}_{4}^{(k+1)}}{\partial\mathbf{D}_{\mathbf{\Phi}_{\tau}}}\right]^{\mathrm{H}}\frac{\partial\mathbf{D}_{\mathbf{\Phi}_{\tau}}}{\partial[\mathbf{\Phi}_{\tau}]_{i,m}}\right\}\nonumber\\
&= \text{tr}\!\left\{\!\mathbf{\Lambda}^{\mathrm{H}}\!\left(\mathbf{R}_{g'_m}\!\otimes\!\frac{\mathbf{R}_c\left(\partial\mathbf{\Phi}_{\tau}\right)^{\mathrm{H}}}{\partial[\mathbf{\Phi}_{\tau}]_{i,m}}\right)\!\right\}
\!=\!\text{tr}\!\left\{{\mathbf{\Lambda}}^{\mathrm{H}}\!\left(\mathbf{R}_{g'_m}\!\otimes\!\mathbf{R}_{c}\mathbf{J}^{\mathrm{H}}_{i,m}\right)\!\right\}.
\end{align}
Then, letting $z =(m-1)\tau+i$, the $z$-th element of the  gradient of \eqref{MSE_k_lambda} with respect to $\mathbf{\Phi}_{\tau}$ is computed as
\vspace{-0.1cm}
\begin{align}\label{Gradient_2}
&\left[\nabla_{\boldsymbol{\Phi}_{\tau}}\mathcal{E}_{x}\left(\mathbf{\Lambda},\mathbf{\Phi}_{\tau}\right) \right]_{z}= \nonumber\\
&\text{tr}\left\{\boldsymbol{\Lambda}^{\mathrm{H}} \boldsymbol{\Lambda}
\left[\mathbf{R}_{g'_m}\otimes\left(\mathbf{J}_{i,m}\mathbf{R}_{c}\mathbf{\Phi}_{\tau}^{\mathrm{H}}+\mathbf{\Phi}_{\tau}\mathbf{R}_{c}\mathbf{J}_{i,m}^{\mathrm{H}}\right)\right]\right\}+\nonumber\\
&\frac{\sigma_{e}^2}{\rho^{\mathrm{tr}}}\text{tr}\left\{{\boldsymbol{\Lambda}}^{\mathrm{H}} \boldsymbol{\Lambda}
\left[\mathbf{R}_{g'_m}\otimes\left(\left(\mathbf{J}_{i,m}\mathbf{R}_{q}\mathbf{\Phi}_{\tau}^{\mathrm{H}}+\mathbf{\Phi}_{\tau}\mathbf{R}_{q}\mathbf{J}_{i,m}^{\mathrm{H}}\right)\odot\mathbf{I}_{\tau}\right)\right]\right\}-\nonumber\\
&\text{tr}\left\{\boldsymbol{\Lambda}\!\left(\mathbf{R}_{g'_m}\otimes\mathbf{J}_{i,m}\mathbf{R}_{c}\right)\!\right\}\!-\!
\text{tr}\left\{\boldsymbol{\Lambda}^{\mathrm{H}}\!\left(\mathbf{R}_{g'_m}\otimes\mathbf{R}_{c}\mathbf{J}_{i,m}^{\mathrm{H}}\right)\!\right\},
\end{align}
where $\mathbf{J}_{i,m}$ is a $\tau \times M$ matrix with the $(i,m)$-th element being $1$ and the other elements being $0$.

As in the case of the gradient operator, letting $t =(s-1)Q+r$ and $z =(q-1)Q+p$, the complex \emph{Hessian} operator $\nabla^{2}_{\mathbf{X}}f$ is the $PQ\times PQ$-dimensional Hermitian matrix defined as 
\begin{equation}
\left[\nabla^{2}_{\mathbf{X}}f \right]_{t,z}= \frac{\partial^{2} f}{\partial \mathbf{x}_{t}^{*}\partial \mathbf{x}_{z}}= \frac{\partial^{2} f}{\partial \mathbf{X}_{r,s}^{*}\partial \mathbf{X}_{p,q}}.
\end{equation}
After that, taking derivatives of \eqref{g_E_1} and \eqref{g_E_2} with respect to $[\mathbf{\Phi}_{\tau}]_{i,m}$ again as follows: 
\begin{align} \label{H_E_1}
&\frac{\partial\nabla_{[\mathbf{\Phi}_{\tau}]_{i_1,m_1}}\mathcal{E}_{1}^{(k+1)}}{\partial[\mathbf{\Phi}_{\tau}]_{i_2,m_2}}  \nonumber\\
&= \text{tr}\left\{{\mathbf{\Lambda}}^{\mathrm{H}}\mathbf{\Lambda}\!\left[\mathbf{R}_{g'_m}\!\otimes\!\frac{\mathbf{J}_{i_1,m_1}\mathbf{R}_c\left(\partial\mathbf{\Phi}_{\tau}\right)^{\mathrm{H}}\!+\!\left(\partial\mathbf{\Phi}_{\tau}\right)\mathbf{R}_c\mathbf{J}_{i_1,m_1}^{\mathrm{H}}}{\partial[\mathbf{\Phi}_{\tau}]_{i_2,m_2}}\!\right]\!\right\}\nonumber\\
&=\text{tr}\left\{\boldsymbol{\Lambda}^{\mathrm{H}} \boldsymbol{\Lambda}\left[\mathbf{R}_{g'_m}\otimes\left(\mathbf{J}_{i_1,m_1}\mathbf{R}_{c}\mathbf{J}_{i_2,m_2}^{\mathrm{H}}+\mathbf{J}_{i_2,m_2}\mathbf{R}_{c}\mathbf{J}_{i_1,m_1}^{\mathrm{H}}\right)\right]\right\},
\end{align}
and
\begin{align} \label{H_E_2}
&\frac{\partial\nabla_{[\mathbf{\Phi}_{\tau}]_{i_1,m_1}}\mathcal{E}_{2}^{(k+1)}}{\partial[\mathbf{\Phi}_{\tau}]_{i_2,m_2}} = \frac{\sigma_{e}^2}{\rho^{\mathrm{tr}}}\text{tr}\left\{{\mathbf{\Lambda}}^{\mathrm{H}}\mathbf{\Lambda}\cdot\right.\nonumber\\
&\left.\left[\mathbf{R}_{g'_m}\otimes\left(\frac{\mathbf{J}_{i_1,m_1}\mathbf{R}_q\left(\partial\mathbf{\Phi}_{\tau}\right)^{\mathrm{H}}+\left(\partial\mathbf{\Phi}_{\tau}\right)\mathbf{R}_q\mathbf{J}_{i_1,m_1}^{\mathrm{H}}}{\partial[\mathbf{\Phi}_{\tau}]_{i_2,m_2}}\!\right)\!\odot\!\mathbf{I}_{\tau}\!\right]\!\right\}\nonumber\\
&=\frac{\sigma_{e}^2}{\rho^{\mathrm{tr}}}\cdot\nonumber\\
&\text{tr}\!\left\{\!\boldsymbol{\Lambda}^{\mathrm{H}}\! \boldsymbol{\Lambda}\!\left[\mathbf{R}_{g'_m}\!\otimes\!\left(\!\left(\mathbf{J}_{i_1,m_1}\mathbf{R}_{q}\mathbf{J}_{i_2,m_2}^{\mathrm{H}}\!\!+\!\mathbf{J}_{i_2,m_2}\mathbf{R}_{q}\mathbf{J}_{i_1,m_1}^{\mathrm{H}}\right)\!\odot\!\mathbf{I}_{\tau}\!\right)\!\right]\!\right\},
\end{align}
and introducing  $z_{1} = (m_{1}-1)\tau+i_1$ and $z_{2} = (m_{2}-1)\tau+i_2$, the Hessian  of \eqref{MSE_k_lambda} with respect to $\mathbf{\Phi}_{\tau}$ is computed as.
\begin{equation}\label{Hessian_2}
\begin{aligned}
&\left[\nabla^{2}_{\boldsymbol{\Phi}_{\tau}}\mathcal{E}_{x}\left(\mathbf{\Lambda},\mathbf{\Phi}_{\tau}\right) \right]_{z_1,z_{2}} =\frac{\sigma_{e}^2}{\rho^{\mathrm{tr}}}\cdot\\
&\text{tr}\!\left\{\!\boldsymbol{\Lambda}^{\mathrm{H}}\! \boldsymbol{\Lambda}\!\left[\mathbf{R}_{g'_m}\!\otimes\!\left(\!\left(\mathbf{J}_{i_1,m_1}\mathbf{R}_{q}\mathbf{J}_{i_2,m_2}^{\mathrm{H}}\!\!+\!\mathbf{J}_{i_2,m_2}\mathbf{R}_{q}\mathbf{J}_{i_1,m_1}^{\mathrm{H}}\right)\!\odot\!\mathbf{I}_{\tau}\!\right)\!\right]\!\right\}\\
&+\text{tr}\left\{\boldsymbol{\Lambda}^{\mathrm{H}} \boldsymbol{\Lambda}\left[\mathbf{R}_{g'_m}\otimes\left(\mathbf{J}_{i_1,m_1}\mathbf{R}_{c}\mathbf{J}_{i_2,m_2}^{\mathrm{H}}+\mathbf{J}_{i_2,m_2}\mathbf{R}_{c}\mathbf{J}_{i_1,m_1}^{\mathrm{H}}\right)\right]\right\}.
\end{aligned}
\end{equation}

Moreover, considering the special scenario of SISO, in this case, $\mathbf{R}_{x}$ and $\mathbf{R}^{\mathrm{tr}}_{w}$ degenerate to $\mathbf{R}_{c}$ and $\left(\!\left(\mathbf{\Phi}_{\tau}\mathbf{R}_{q}\mathbf{\Phi}_{\tau}^{\mathrm{H}}\right)\odot\mathbf{I}_{\tau}\right)$. Therefore, the gradient and Hessian of \eqref{MSE_k_lambda} with respect to $\mathbf{\Phi}_{\tau}$ can be simplified as
\begin{align}\label{Gradient_siso_ch}
&\nabla_{\boldsymbol{\Phi}_{\tau}}\mathcal{E}_{x}\left(\mathbf{\Lambda},\mathbf{\Phi}_{\tau}\right) =\nonumber\\
& 2\boldsymbol{\Lambda}^{\mathrm{H}}\boldsymbol{\Lambda} \boldsymbol{\Phi}_{\tau} \mathbf{R}_{c} 
+ \frac{2\sigma_{e}^2}{\rho^{\mathrm{tr}}}\left(\!\left(\boldsymbol{\Lambda}^{\mathrm{H}}\boldsymbol{\Lambda}\right)\odot\mathbf{I}_{\tau}\right) \boldsymbol{\Phi}_{\tau}\mathbf{R}_{q} - 2 {{}\boldsymbol{\Lambda}}^{\mathrm{H}} \mathbf{R}_{c},
\end{align}
and
\begin{align}\label{Hessian_siso_ch}  
\nabla^{2}_{\boldsymbol{\Phi}_{\tau}}\mathcal{E}_{x}\left(\mathbf{\Lambda},\mathbf{\Phi}_{\tau}\right) = 2\boldsymbol{\Lambda}^{\mathrm{H}}\boldsymbol{\Lambda} \otimes \mathbf{R}_c + \frac{2\sigma_{e}^2}{\rho^{\mathrm{tr}}}\left(\!\left(\boldsymbol{\Lambda}^{\mathrm{H}}\boldsymbol{\Lambda}\right)\odot\mathbf{I}_{\tau}\right) \otimes \mathbf{R}_q.
\end{align}

Similarly, during the data transmission phase, abbreviating ${\mathbf{v}}^{(k+1)}$ as $\mathbf{v}$, the elements of the gradient and Hessian of \eqref{MSE_s11} with respect to $\boldsymbol{\phi}$ are computed as
\begin{align}\label{G_MSE_2_com}
&\left[\nabla_{\boldsymbol\phi}\mathcal{E}_{s}\left(\mathbf{v},\boldsymbol\phi;\hat{\mathbf{x}}\right) \right]_{m} =  -2\sqrt{\rho}\text{tr}\left\{\hat{\mathbf{x}}\mathbf{v}^{\mathrm{H}}\mathbf{J}_{Nm}\right\}\nonumber\\
&+\rho\text{tr}\left\{{\mathbf{v}}\mathbf{v}^{\mathrm{H}}
\left[\mathbf{J}_{Nm}\left( \hat{\mathbf{x}}\hat{\mathbf{x}}^{\mathrm{H}}+\mathbf{R}_{\tilde x}+ \frac{\sigma_e^2}{\rho}\left(\mathbf{R}_{g'_m}\otimes\mathbf{R}_{q}\right)\right){\boldsymbol\Phi}_N^{\mathrm{H}}\right.\right.\nonumber\\
&+\left.\left.{\boldsymbol\Phi}_N\left(\hat{\mathbf{x}}\hat{\mathbf{x}}^{\mathrm{H}}+\mathbf{R}_{\tilde x}+ \frac{\sigma_e^2}{\rho}\left(\mathbf{R}_{g'_m}\otimes\mathbf{R}_{q}\right)\right)\mathbf{J}_{Nm}^{\mathrm{H}}\right]\right\},
\end{align}
and
\begin{align} \label{Hessian_s_2}
&\left[\nabla^2_{\boldsymbol\phi}\mathcal{E}_{s}\left(\mathbf{v},{\boldsymbol\phi};\hat{\mathbf{x}}\right) \right]_{m_1,,m_2} = \nonumber\\
&\rho\text{tr}\left\{\mathbf{v}{\mathbf{v}}^{\mathrm{H}}
\left[\mathbf{J}_{N m_1}\left(\hat{\mathbf{x}}\hat{\mathbf{x}}^{\mathrm{H}}+\mathbf{R}_{\tilde x}+ \frac{\sigma_e^2}{\rho}\left(\mathbf{R}_{g'_m}\otimes\mathbf{R}_{q}\right)\right)\mathbf{J}_{Nm_2}^{\mathrm{H}}\right.\right.\nonumber\\
&+\left.\left.\mathbf{J}_{N m_2}\left(\hat{\mathbf{x}}\hat{\mathbf{x}}^{\mathrm{H}}+\mathbf{R}_{\tilde x}+ \frac{\sigma_e^2}{\rho}\left(\mathbf{R}_{g'_m}\otimes\mathbf{R}_{q}\right)\right)\mathbf{J}_{N m_1}^{\mathrm{H}}\right]\right\},
\end{align}
where $\mathbf{J}_{Nm}=\mathbf{I}_{N}\otimes\mathbf{J}_{m}$ and $\mathbf{J}_{m}$ is an $M$-dimensional vector with the $m$-th element being
1 and other elements being 0. Correspondingly, in the SISO scenario, where $v$ degenerates into a complex number, the gradient and Hessian can be written as
\begin{align}\label{Gradient_siso_com}
&\nabla_{\boldsymbol\phi}\mathcal{E}_{s}\left(v,\boldsymbol\phi;\hat{\mathbf{x}}\right) = \nonumber\\
&2\rho\left|v\right|^2\boldsymbol{\phi}\left(\hat{\mathbf{x}}\hat{\mathbf{x}}^{\mathrm{H}}+\mathbf{R}_{\tilde x}+ \frac{\sigma_e^2}{\rho}{\mathbf{R}_q}\right) 
- 2\sqrt{\rho}v^*{\hat{\mathbf{x}}}^{\mathrm{H}},
\end{align}
and
\begin{align}\label{Hessian_siso_com}  
\nabla^2_{\boldsymbol\phi}\mathcal{E}_{s}\left(v,\boldsymbol\phi;\hat{\mathbf{x}}\right) = 2\rho\left|v\right|^2\left(\hat{\mathbf{x}}\hat{\mathbf{x}}^{\mathrm{H}}+\mathbf{R}_{\tilde x}+ \frac{\sigma_e^2}{\rho}{\mathbf{R}_q}\right).
\end{align}


\bibliographystyle{IEEEtran}
\bibliography{IEEEabrv,reference}

\end{document}